# FROM LOCAL TO GLOBAL:
# EXTERNAL VALIDITY IN A FERTILITY NATURAL EXPERIMENT

Rajeev Dehejia, Cristian Pop-Eleches, and Cyrus Samii[*]

December 2018

*Forthcoming at Journal of Business and Economic Statistics*

[*]Dehejia, Wagner Graduate School of Public Service, New York University, 295 Lafayette Street, New York, NY 10012 (Email: rajeev@dehejia.net). Pop-Eleches, School of International and Public Affairs, Columbia University, 420 W 118th Street, New York, NY 10027, (Email: cp2124@columbia.edu). Samii, Department of Politics, New York University, 19 West 4th Street, New York, NY 10012 (Email: cds2083@nyu.edu). The authors thank Morris Chow for excellent research assistance; Ali T. Ahmed, Hunt Allcott, Joshua Angrist, Peter Aronow, Neal Beck, James Bisbee, Gary Chamberlain, Drew Dimmery, Michael Gechter, Rachel Glennerster, and Raimundo Undurraga for valuable comments and suggestions; and seminar participants at the BREAD conference, Cowles Econometrics Seminar, EGAP, the Federal Reserve Board of Cleveland, the Federal Reserve Board of New York, Georgetown, GREQAM, IZA, Maastricht, NEUDC 2014, Columbia, Harvard, MIT, NYU, UCLA, UCSD, the World Bank, Yale, the 2014 Stata Texas Empirical Microeconomics Conference, and the Stanford 2015 SITE conference for helpful feedback.

# FROM LOCAL TO GLOBAL:
# EXTERNAL VALIDITY IN A FERTILITY NATURAL EXPERIMENT

**Abstract**

We study issues related to external validity for treatment effects using over 100 replications of the Angrist and Evans (1998) natural experiment on the effects of sibling sex composition on fertility and labor supply. The replications are based on census data from around the world going back to 1960. We decompose sources of error in predicting treatment effects in external contexts in terms of macro and micro sources of variation. In our empirical setting, we find that macro covariates dominate over micro covariates for reducing errors in predicting treatments, an issue that past studies of external validity have been unable to evaluate. We develop methods for two applications to evidence-based decision-making, including determining where to locate an experiment and whether policy-makers should commission new experiments or rely on an existing evidence base for making a policy decision.

# 1. Introduction

In recent decades across a wide range of fields in economics, such as labor, education, development, and health, the use of experimental and quasi-experimental methods has become widespread. The emphasis on experimental and quasi-experimental methods[1] was driven by an attempt to generate internally valid results. At the same time, the global scale of experiments points to the less-emphasized but central concern of external validity. In evaluating the external validity of a set of experiments, one poses the question, "to what population, settings, and variables can this effect be generalized?" (Campbell 1957). In other words, external validity can be measured in terms of the error in prediction of treatment effects for new populations beyond those covered in the evidence base. With a single or handful of studies in a limited range of contexts, external validity is mostly a matter of theoretical speculation. But with a large number of internally valid studies across a variety of contexts, it is reasonable to hope that researchers are accumulating generalizable knowledge, i.e., not just learning about the specific time and place in which a study was run but about what would happen if a similar intervention were implemented in another time or place.

The success of an empirical research program can be judged by the diversity of settings in which a treatment effect can be reliably predicted, possibly obviating the need for further experimentation with that particular treatment. This is the issue we address in this paper. More specifically, given internally valid evidence from "reference" settings, is it possible to predict the treatment effect in a new ("target") setting? Is it possible to understand how differences between actual and predicted treatment effects vary with differences between the setting of interest and the settings in which experimental evidence is available? And if so which differences are more important: context-level (e.g., macro or institutional) variables or individual-level micro variables? How might we judge whether an existing evidence base is adequate for informing new policies, thereby making further experiments with a given treatment unnecessary?

---
[1] Throughout the remainder of the paper, we will use the term *experiments* broadly as referring to internally valid studies that use either true random experimental or quasi-experimental methods.

Although the issue of external validity has garnered the most attention recently in the context of randomized controlled trials, it is important to underline that the essential challenge of extrapolation is common to the broad set of methods used to identify treatment effects. Each of these methods has its own specific challenges for extrapolation. In this paper, as a starting point, we focus on reduced-form experiments or natural experiments. In ongoing and future work, we extend the analysis to other research designs (see for example Bisbee, Dehejia, Pop-Eleches, and Samii 2017 for a related analysis of the instrumental variables case).

Our approach in this paper is to use a natural experiment for which "replications" are, in fact, available for a wide variety of settings. We use the Angrist and Evans's (1998) research design that treats sex-composition (same sex of the first two children) as exogenous to define a natural experiment with outcomes being incremental fertility (having a third child) and mother's labor supply. Replications of this natural experiment are recorded for a large number of countries over many years in censuses compiled in the Integrated Public Use Microdata Series - International (IPUMS-I) data. Cruces and Galiani (2007) and Ebenstein (2009) have studied how the effects in this natural experiment generalize to Argentina and Mexico and to Taiwan, respectively. Our analysis extends this to all available IPUMS-I samples around the world going back to 1960, allowing for a very rich examination of both micro- and macro-level sources of heterogeneity. Filmer, Friedman, and Schady (2009) estimate effects of sex composition on incremental fertility (but not labor supply) for mothers in different regions around the world. Compared to our approach, their primary focus is on understanding son-preferred differential fertility-stopping behavior and since they are using Demographic and Health Survey data, their samples tend to over represent developing countries. Their results show that the effect of sex composition on incremental fertility is apparent around the world, particularly in trying to make up for the absence of sons in early births.

We discuss the strengths and weaknesses of our data in greater detail in Section 4. But, briefly, it is important to acknowledge that *Same-Sex* is not a perfect natural experiment when estimated on a global scale. To the extent that fertility choices could be viewed as culture- and context-specific, we believe we are setting a high bar for the exercise: if we are able to find a degree of external validity for a fertility natural experiment, then there is hope that it might be possible for other experiments as well.

The paper is both a methodological "thought experiment" and an empirical investigation. As a thought experiment, we consider the rather fanciful situation of having replications of an experiment or well-identified result across a wide variety of contexts that we can use to inform an extrapolation to an external setting. This is an idealized setting in certain respects, given the large number of sites and also the homogeneity in treatments and outcomes. What brings us back down to earth is that we have only a limited amount of information that we can use to characterize effect heterogeneity. This situation applies to many empirical studies. As an empirical investigation, our task is to assess the external validity potential of this evidence base in extrapolating to new contexts. The evidence base consists of the set of studies and its limitations are defined by the variety of contexts that it covers and, crucially, the measured covariates that it includes. We approach the extrapolation problem as empiricists, using the available data in an agnostic and flexible manner. Our application is especially conducive to such an agnostic approach, because we have many contexts, large within-context sample sizes, and a relatively spare set of micro-level covariates, which allows us to use saturated specifications. In addition, the inferential goal is to predict an effect in a target context that is directly analogous to effects that we can observe in reference contexts. In other settings, analysts may do better to draw from theoretically-informed models. This includes cases where the inferential goal is to predict a counterfactual for which existing experiments provide only indirect information, or cases where theory can inform parametric restrictions that allow for more efficient estimation with modest sample sizes. We examine how working through the extrapolation problem using the evidence base can inform how an experimental or quasi-experimental research program might optimally proceed. A complementary exercise, which we do not undertake in this paper, would be to use the evidence base to explain effect heterogeneity for the sake of theory development (see Aaronson, Dehejia, Jordan, Pop-Eleches, Samii, and Schulze 2017).

The topic of external validity has been gathering increasing attention in the economics literature. Empirical assessments of external validity in economics include recent work by Allcott (2014), Andrews and Oster (2018), Bell et al. (2016), Gechter (2015), Pritchett and Sandefur (2013), and Vivalt (2014). Using two examples from the education literature (class size effects and the gains from private schooling), Pritchett and Sandefur (2013) argue that that economy-wide or institutional characteristics often dominate the importance of individual characteristics when attempting to extrapolate treatment effects. With a large number of (natural) experiments in our data set (over 100 replications compared to the dozen or so studies they use in their analysis) we are able to address this question with evidence that has broad temporal and geographic coverage. (They also argue that estimates from observational studies within a context are superior to extrapolated experimental results from other contexts. We address this question in Bisbee, Dehejia, Pop-Eleches, and Samii 2017.)

Vivalt (2014) uses a random effects meta-analysis to study sources of effect heterogeneity for sets of development program impact evaluations. She finds evidence of program effects varying by the implementing actor, with government programs tending to fare worse than non-governmental organization programs. She also finds that with a small set of study-level characteristics (namely, implementer, region, intervention type, and outcome type), meta-regressions have only modest predictive power. In our analysis, we consider a somewhat larger number of covariates both at the micro- and macro-levels and we do so in a set of experiments that is more homogenous in terms of treatments and outcomes. This allows us to distinguish issues of extrapolation from questions of outcome and treatment comparability.

Our results show that there is considerable treatment effect heterogeneity in the effect of sex composition on fertility and labor supply across country-years, but that some of this variation can be meaningfully explained both by individual and context (experiment – in our case country-year – level) covariates. We define and estimate an "external validity function" that characterizes the quality of an evidence base's predictions for a target setting. We examine the relationship between prediction error and individual and context covariates. While both are potentially useful in reducing prediction error from external comparisons, in our application context variables dominate. This is in part a feature of our set up---namely, the nature of the effects in our application, the sparse set of micro covariates that we have at our disposal, and also the fact that some of our context covariates (e.g., aggregate labor force participation rates) are closely related to the micro-level outcomes. But it is an important finding nonetheless, because methodological work on treatment effect extrapolation (reviewed below) has tended to focus on accounting for variation in micro-level variables. Moreover, we find that context interaction effects are important, such that the effects of the micro-level variables tend to depend on context variables (e.g., a 35-year-old woman in a lower income country may have different potential outcomes than a woman of the same age in a high-income country). Our analysis and empirical results indicate the need to take context-level heterogeneity into consideration for extrapolating treatment effects.

Finally, we present two applications to evidence-based decision-making. In the first, we use the external validity function to determine the best location of a new experiment. Specifically, choosing among our country-year sites, we ask which location would minimize mean squared prediction error for the other sites? In the second application, we ask when a policy decision maker should choose to run an experiment in a target setting rather than use extrapolated estimates of the treatment effect from an existing evidence base. For both applications, pre-treatment covariate data proves to be crucial. Questions of external validity motivate the collection of rich covariate data even when an experiment or natural experiment does not require it for internal validity.

The paper is organized as follows. In Section 2, we provide a brief review of the related literature, while in Section 3 we outline a simple analytic framework for our empirical analysis. In Section 4, we discuss our data and the sex composition natural experiment. In Section 5 we present a graphical analysis of treatment effect heterogeneity, and in Section 6 we perform the analogous hypothesis tests to reject homogenous treatment effects. In Section 7, we present non-parametric estimates of the external validity function for selected covariates of interest. In Section 8, we use multivariate regressions to examine the relative importance of individual and context-level predictors in determining the external validity of experimental evidence. In Section 9, we present evidence on the out-of-sample predictive accuracy of the model, and in particular examine how external validity evolves with the accumulation of evidence. Section 10 presents our two applications, the choice of experimental site and of whether or not to run an experiment to inform a policy decision. Section 11 concludes.

**2. Related methodological literature**

Our analysis follows on the call by Imbens (2010) to scrutinize empirically questions of external validity, rather than relying only on theoretical speculation. Focused consideration of external validity goes back at least to Campbell (1957), whose approach is taken up by Shadish et al. (2002). Debates in the classical literature omit a formal statement of how external validity may be achieved. Olsen et al. (2013) derive an expression for "external validity bias" that characterizes how a treatment effect estimate from a subset of contexts may differ from the treatment effect in the complete set of contexts. We pursue a different approach based on treatment effect extrapolation. This follows recent work by Hotz, Imbens, and Mortimer (2005), Stuart et al. (2011), Hartman et al. (2015), and Tipton (2013), all of whom use the potential outcomes framework to characterize conditions necessary for extrapolation from a reference population for which experiments are available to a target population. These conditions are analogous to those required for using covariates to identify causal effects under "strong ignorability" (Rosenbaum and Rubin, 1983). The difference is that the relevant conditional independence assumptions pertain to inclusion in the reference versus target population rather than in the treatment versus control group. Making use of such identifying conditions requires measuring statistical relations between covariates and treatment effects that are invariant as we move from the reference to the target population (Heckman and Vytlacil, 2007; Pearl and Bareinboim, 2014). We review these conditions in the next section.

Hotz et al. (2005), Cole and Stuart (2010), Stuart et al. (2011), Hartman et al. (2015), and Kern et al (2016) apply various approaches to extrapolation from one site to another, including matching, inverse probability weighting, and parametric and non-parametric regression techniques. Crump et al. (2008), Green and Kern (2012), and Imai and Ratkovic (2013) develop non-parametric methods for characterizing effect heterogeneity, including sieve estimators, Bayesian additive regression trees, and support vector machines, respectively. Because these previous studies work with only a small number of sites, they focus on micro-level differences across sites. Our analysis addresses both micro-level differences and macro-level differences (that is, country-year-level contextual characteristics). In a recent study that comes closer to what we do here, Orr et al. (2017) work with the results of multisite education experiments, using a leave-one-out approach to examine the out-of-sample predictive accuracy of multilevel mixed-effects regression models that model treatment effects as functions of context-level variables. In a similar spirit, Bloom et al. (2016) use a multilevel mixed-effects regression model to estimate the variance of treatment effects across sites. Angrist (2004), Angrist and Fernandez-Val (2010), and Aronow and Sovey (2013) consider extrapolation from local average treatment effects identified by instrumental variables to a target population. We avoid this issue in the current discussion, as we focus only on reduced form or intention-to-treat effects. We address extrapolation with instrumental variables in related work (Bisbee, Dehejia, Pop-Eleches, and Samii 2017).

Our analysis is related to the meta-analysis literature (Glass, 1976; Hedges and Olkin, 1985; Sutton and Higgins, 2008). Applications in economics include Bloom et al. (2003), Card et al. (2010), Dehejia (2003), and Stanley (2001), as well as meta-analytic reviews that appear in the *Journal of Economic Surveys*. What the meta-analysis literature lacks, however, is a general (i.e., non-parametric) characterization of the conditions required for consistent extrapolation from reference to target populations. Classical approaches to meta-analysis use meta-regression to determine correlates of effect heterogeneity---so called "moderator" analysis. The classical literature tends to leave unclear the purpose of such moderator analysis with some discussions suggesting that it is merely descriptive, with no claim of identifying an effect in a target population, and others suggesting the much more ambitious goal of trying to establish a full generative model of the conditional effect distribution (Greenland 1994; Rubin 1992). The work on non-parametric identification of extrapolated effects, which we use as the foundation of our analysis, is explicit about conditions for either identifying moderator effects or consistent extrapolation to new populations.

**3. Analytical framework**

We have a set of $C$ contexts, indexed by $c = 1, \dots, C$, drawn from some global population of contexts. In applied settings, the set of contexts may be sampled in a manner that is not completely at random, and so our analysis does not take this for granted. Rather, as we specify formally below, what we need for consistent extrapolation is for units across contexts to be exchangeable conditional on covariates. Thus, each context is characterized by a vector of context-level covariates, $V_c$. Within each sampled context we have $N_c$ units indexed as $i = 1, \ldots, N_c$, drawn from the context's population of units. Each of these units is characterized by a vector of unit-level covariates, $W_{ic}$. Our interest is in causal effects for a unit-level binary treatment, $T_{ic} \in \{0,1\}$. Each unit $i$ in context $c$ is characterized by a pair of potential outcomes, $Y_{ic}(0)$ for the outcome under the control condition ($T_{ic} = 0$), and $Y_{ic}(1)$ for the outcome under the treatment condition ($T_{ic} = 1$).

We consider a data generating process in which individuals from one of contexts are selected as the targets for which we want to predict the average treatment effect, and the individuals in the other $C - 1$ contexts serve as reference cases to use in formulating these predictions. We refer to the context that contains the target units as the "target context" and the contexts containing the reference units as the "reference contexts." The set up is similar to that of Hotz et al. (2005), except that we consider situations with potentially many reference contexts, and so adjusting for context-level variables is a practical possibility. This corresponds to our empirical setting, in which available reference experiments accumulate over time.

To formalize this selection process, suppose that each unit is also characterized by an indicator variable $D_{ic} \in \{0,1\}$ for whether a unit is a member of the target population or from a reference context. Members of the target population have $D_{ic} = 1$, and units residing in one of the $C - 1$ reference contexts have $D_{ic} = 0$. In the reference contexts, experiments are run that randomly assign the treatment ($T_{ic}$) to sampled units, revealing outcomes as

(1) $\quad Y_{ic} = T_{ic} Y_{ic}(1) + (1 - T_{ic}) Y_{ic}(0).$

Note that expression (1) embeds the "stable unit treatment value assumption" (SUTVA; Rubin 1980). For units in the reference contexts, we observe $(Y_{ic}, T_{ic}, W_{ic}, D_{ic}, V_c)$, and for target units, we observe only $(W_{ic}, D_{ic}, V_c)$. Note that we are assuming that we have access to micro-level data in both the reference and target contexts, and the relevance of this assumption depends on whether the substance of the enquiry is one for which relevant survey or census data is available.

Suppose the following conditions on the data generating process:

(C0)     $T_{ic} \perp\!\!\!\perp (Y_{ic}(0), Y_{ic}(1)) | (V_c, W_{ic}), D_{ic} = 0$,

(C1)     $D_{ic} \perp\!\!\!\perp (Y_{ic}(0), Y_{ic}(1)) | (V_c, W_{ic})$, and

(C2)     $\delta < \Pr[D_{ic} = 0 | V_c = v, W_{ic} = w] < 1 - \delta$ for $\delta > 0$ and all $(v, w)$ in the support of $(V, W)$.

Condition C0 means that in the reference contexts we have random assignment with respect to potential outcomes, conditional on covariates. This implies that conditional treatment effects are identified in each of the reference contexts. Condition C1 requires that systematic differences in outcomes across target units and units in the reference contexts depend only on $V_c$ and $W_{ic}$. Condition C2 means that for all covariates values, one can expect to find units in the samples from reference contexts. If C2 is not satisfied unconditionally, one can redefine the target population as being the sub-population for which common support holds (Hotz et al. 2005, fn. 7). Conditions C1 and C2 mean that conditional and average treatment effects for target units are identified from the units in the reference contexts by conditioning on covariates.

Our estimand is the average treatment effect for target units. Define $E[A]$ as the expected value of $A$ given the distribution induced by sampling, selection of target units, and treatment assignment, and define the conditional expectation $E[A|B = b]$ similarly for the distribution of $A$ in the subset of units for which $B = b$. Then, our estimand is

(2) $\quad \tau_1 = E[Y_{ic}(1) - Y_{ic}(0)|D_{ic} = 1]$ .

As per Hotz et al. (2005, Lemma 1), C0-C2 imply that $\tau_1$ is identified from the data in the reference contexts:

$$
\begin{aligned}
(3) \quad \tau_1 &= E[E[Y_{ic}(1) - Y_{ic}(0)|D_{ic} = 1, V_c = v, W_{ic} = w]|D_{ic} = 1] \\
&= E[E[Y_{ic}(1) - Y_{ic}(0)|D_{ic} = 0, V_c = v, W_{ic} = w]|D_{ic} = 1] \\
&= E[E[Y_{ic}|T_{ic} = 1, D_{ic} = 0, V_c = v, W_{ic} = w]|D_{ic} = 1] \\
&\quad - E[E[Y_{ic}|T_{ic} = 0, D_{ic} = 0, V_c = v, W_{ic} = w]|D_{ic} = 1]
\end{aligned}
$$

For the application in section 10.2 below, we specify a conditional mean function as,

(4) $\quad \mu(t, v, w) = E[Y_{ic}|T_{ic} = t, D_{ic} = 0, V_c = v, W_{ic} = w],$ .

for $t = 0,1$. We estimate this conditional mean function using a series regression with polynomial expansions and interactions of the covariates. We fit these models using ordinary least squares, and the order of the polynomials and interactions are determined using minimum-$C_p$ LASSO regularization, as in Belloni et al. (2014). The LASSO regularization helps to identify a series specification with high predictive accuracy but in a manner that reduces the risk of overfitting. These regressions yield a conditional mean estimator, $\hat{\mu}(t, v, w)$ for $t = 0,1$. Define $V_{(1)}$ as the value of the context level covariate that obtains for the target context, $W_{i(1)}$ as the covariate value for target unit $i$, and $S_{(1)}$ as the set of sampled target units. Assuming that the conditional mean estimator is consistent for expression (4), an asymptotically unbiased estimator for $\tau_1$ needs to take the conditional mean estimates and marginalize with respect to the covariate distribution of the target population. We implement this with the following estimator,

(5) $\quad \hat{\tau}_1 = \frac{1}{|S_{(1)}|}\sum_{i \in S_{(1)}}[\hat{\mu}(1, V_{(1)}, W_{i(1)}) - \hat{\mu}(0, V_{(1)}, W_{i(1)})]$ ,

where unbiased marginalization follows from the random sampling of units from the target population. (Note that since we only ever observe one target context covariate distribution, we have asymptotic unbiasedness but not consistency---the error is non-vanishing. We address this below.) This approach is similar to the "response surface modeling" approach of Orr et al. (2017, eq. 4), although in our case, we allow for covariates to moderate effects at both the unit and context levels, whereas Orr et al. only model how covariates moderate context-level effect heterogeneity.

In our analysis below, we compare predictions to the treatment effects that actually arise for each target population. For any given target population, we can call the treatment effect that arises $\tau_{(1)}$. We can relate a given target population's treatment effect to the expected value, $\tau_1$, as

(6) $\quad \tau_{(1)} = \tau_1 + \epsilon_{(1)}.$

The term $\epsilon_{(1)}$ captures what we call "intrinsic variation" in target population treatment effects. It is analogous to the error between a conditional mean and an observation in a regression setting. If assumptions C0-C2 hold, this error is zero in expectation (with respect to the notion of expectations defined above). For any given target, however, the error could be small or large, and conditional on $(V_{(1)}, W_{i(1)})$ it may not be mean zero.

The identification results above are for $\tau_1$ but then if our interest is really in $\tau_{(1)}$, how should we address the issue that the two differ as characterized in expression (6)? Our approach, as developed below (section 10.2) is to construct $(1-\alpha)100\%$ predictive intervals that, under substantive assumptions on the between-context variation in treatment effects, are calibrated to cover the target $\tau_{(1)}$ values with probability $1-\alpha$. Analogous to prediction in standard regression analyses, width of the predictive interval measures the degree of uncertainty, and such uncertainty depends on the number of covariates available (which in turn defines the degree of "residual variance") and then the position of the target context in the covariate space (where distance from the centroid of the covariate space tends to imply for uncertainty).

Now, $\tau_{(1)}$ is a population parameter that is not directly observable. Rather, we our empirical analysis uses the realization of the (natural) experiment in the target context to estimate $\hat{\tau}_{(1)}$. We do this using a reduced form regression with an unbiased specification based on the experimental design. Our analyses below focus on the distribution of the "prediction error," defined as

(7) $\quad \hat{\zeta}_{(1)} = \hat{\tau}_1 - \hat{\tau}_{(1)}.$

If both terms on the right hand side are unbiased for their respective estimands, then this prediction error is zero in expectation. The sampling and random assignment processes imply that $\hat{\tau}_{(1)}$ is statistically independent of $\hat{\tau}_1$, and the within-context distribution is asymptotically normal and centered on $\tau_{(1)}$ (Abadie et al. 2014; Freedman 2008; Lin 2013).

Whether $\hat{\tau}_1$ is consistent for $\tau_1$ depends on whether conditions C0-C2 hold and then whether the conditional mean estimators are consistent. Our series specification for the conditional mean estimators is meant to ensure consistency. Our setting is such that C0 is plausible by design, and the unit-level covariate set is relatively small, in which case C2 is also uncontroversial. What remains in question, then, is C1. Below, we conduct descriptive and regression analyses of the distribution of $\hat{\zeta}_{(1)}$ as a way to assess C1. We study whether mean prediction errors go to zero as we align covariate values across the reference and target contexts. Following Hotz et al. (2005) and Gechter (2015), we also test whether expected $Y_{ic}(0)$ values, conditional on $(V_c, W_{ic})$, line up across reference and target contexts, as such equality is an implication of C1.

Assuming consistent estimation, and with a large number of reference contexts and large within-context sample sizes, the distribution of $\hat{\zeta}_{(1)}$ is dominated by the distribution of $\epsilon_{(1)}$. We only ever select one target context, and so the contribution of the variance $\epsilon_{(1)}$ to the variance of $\hat{\zeta}_{(1)}$ does not diminish as we accumulate more experiments—hence the term "intrinsic variation." This is the same as in the analysis of a regression forecast for a single future observation.

We use both dyadic and cumulative analyses to study prediction error and its relationship to covariate differences between reference and target contexts, which we call the "external validity function" (analogous to the bias function in Heckman, Ichimura, Smith and Todd 1998). In the dyadic analysis, we pair each country-year in our sample to each other country-year, creating approximately 28,000 dyads consisting of hypothetical target and reference country-years pairs. In the cumulative analysis, the reference set includes country-years observed in years prior to that of the target country-year. We note that, in the analyses below, we sometimes allow for previous years in a given country to be used as reference contexts for that country. While this may be favorable to the task of extrapolation in some ways, it is worth keeping in mind that sometimes the within-country data are a decade or more apart, in which case it is not clear that within-country data would dominate more contemporaneous data from elsewhere. (See Aaronson, Dehejia, Jordan, Pop-Eleches, Samii, and Schulze 2017 for a relevant analysis.)

## 4. A global natural experiment

There are two main challenges for assessing methods for extrapolating causal effects. First is to find a randomized intervention or a naturally occurring experiment that has been implemented in a wide range of settings around the world. The second is to find data that are readily available and comparable across the different settings.

For the first challenge, we propose to use sibling sex composition to understand its impact on fertility and labor supply decisions. The starting point of our paper is Angrist and Evans (1998), who show, using census data from 1980 and 1990 in the US, that families have on average a preference to have at least one child of each sex. Since gender is arguably randomly assigned, they propose to use the sibling sex composition of the first two children as an exogenous source of variation to estimate the causal impact of fertility on labor supply decision of the mother.

For the second challenge, we make use of recently available data from the Integrated Public Use Microdata Series-International (IPUMS-I). This project is a major effort to collect and preserve census data from around the world. One important dimension of IPUMS-I is their attempt to harmonize the data and variables in order to make them comparable both across time and space. For our application, we work with 142 country-year samples (from 61 unique countries) with information on fertility outcomes as well as country-level covariates, and then 128 country-year samples with data on both fertility and labor-supply decisions as well as country-level covariates.

The use of the Angrist-Evans same-sex experiment on a global scale brings additional challenges, which were not faced in the original paper. In particular, sex selection for the first two births, which does not appear to be a significant factor in the United States (Angrist and Evans 1998), could be a factor in countries where son-preference is a stronger factor than the US. We view sex selectivity as one of the context covariates, *W*, that could be controlled for when comparing experimental results to a new context of interest, or if not appropriately controlled for could undermine external validity. In our results below we pursue three approaches: not controlling for differences in sex selectivity and examining whether external validity still holds; directly examining its effect on external validity; and excluding countries in which selection is known to be widely practiced.

Another challenge is that, if the cost of children depends on sibling sex composition, then the variable *Same-Sex* (which equals 1 if the first two births are the same sex, and zero otherwise) would violate the exclusion restriction that formed the basis of Angrist and Evans's original instrumental variables approach, affecting fertility not only through the taste for a gender balance but also through the cost of additional children (e.g., with two same sex children hand-me-downs lower the cost of a third child and thus could affect not only fertility but also labor supply). Butikofer (2011) examines this effect for a range of developed and developing countries, and argues that this is a concern for the latter group. As a result, in this analysis, we use *Same-Sex* as a reduced-form natural experiment on incremental fertility and on labor supply, and do not present instrumental variables estimates (see Bisbee, Dehejia, Pop-Eleches, and Samii 2017 for an effort to extrapolate the instrumental variables results).

For our empirical analysis, we implement essentially the same sample restrictions, data definitions, and regression specifications as those proposed in Angrist and Evans (1998).[2] Since the census data that we use does not contain retrospective birth histories, we match children to mothers as proposed by Angrist and Evans (1998), using the harmonized relationship codes available through IPUMS-I, and we also restrict our analysis to married women aged 21-35 whose oldest child was less than 18 at the time of the census. In our analysis we define the variable *Same-Sex* to be equal to 1 using the sex of the oldest two children.

---

[2] The data and programs used in Angrist and Evans (1998) are available at:
http://economics.mit.edu/faculty/angrist/data1/data/angev98

As outcomes we use an indicator for the mother having more than 2 children (*Had more children*) and for the mother working (*Economically active*). These two outcomes correspond to the first stage and reduced-form specifications of Angrist and Evans. While there is a natural link between *Same-sex* and *Had more children*, the link is less intuitive for *Economically active*. In the context of instrumental variables, the link is presumably through incremental fertility (and is assumed exclusively to be so). In our application, since no exclusion restriction is assumed, the effect can include not only incremental fertility but also, for example, the income and time effects of having two children of the same sex. As such, identification of the reduced-form effect of *Same-sex* on *Economically active* relies only on the validity of the experiment within each country-year (assumption C0 from Section 3). As we will see below, the contrast between the two reduced form experiments is useful in thinking through issues of external validity.

Next we discuss the choice of individual (micro) and context (macro) variables to be included in our analysis. In the absence of a well-defined theory for our specific context, the choice of individual level variables to explain effect heterogeneity is based on related models and empirical work (Angrist and Evans 1998; Ebenstein 2009). We use the education level of both the mother and the spouse, the age of the mother as well as the age at first marriage for the mother as our main individual level variables. For context variables, obvious candidates are female labor force participation as a broad measure of employment opportunities for women in a given country (Blau and Kahn, 2003) and the total fertility rate. Since the goal of our exercise is extrapolation, we also include a number of macro variables that do not necessarily play a direct causal role in explaining fertility and labor supply decisions but rather have been shown to be important in explaining broad patterns of socio-economic outcomes across countries; these include log GDP per capita, as a broad indicator of development, average education, and geographic distance between reference and target country (Gallup, Mellinger and Sachs, 1999). An important caution is that a number of context variables, but especially labor force participation and fertility, are potentially endogenous at the macro level. In principle, such endogeneity would tend to increase the explanatory power of these variables for explaining effect heterogeneity. But the effects in our applies setting tend to be quite small, and so these variables' explanatory power would likely derive primarily from the fact that they track baseline levels of the outcomes of interest.

Descriptive statistics for our samples are provided in Table 1. On average 60% of women have more than 2 children (*Had more than two kids),* which is our main fertility outcome. Furthermore, 46% of women in our sample report being *Economically active*, which is our main labor market outcome. Summary statistics for a number of additional individual level variables as well as country level indicators are also presented in Table 1 and they include the education of the woman and her spouse, age, age at first birth, and then at the country level, real GDP per capita as well as mean levels for the individual level covariates. We also display summary statistics for the difference in rates of boys versus girls in women's first two births.

For our main empirical specification for each country-year sample, we examine the treatment effect of the *Same-Sex* indicator on two outcome variables (*Had more*

Table 1: Summary Statistics

|  | Mean | S.D. | Obs |
|---|---|---|---|
| *Panel A: Individual level variables* | | | |
| Had more than two kids | 0.60 | 0.49 | 11,766,586 |
| Economically active | 0.46 | 0.50 | 10,275,779 |
| First two children are same sex | 0.51 | 0.50 | 11,766,586 |
| Age | 30.00 | 3.60 | 11,766,586 |
| Education (own) | 1.88 | 0.85 | 11,295,065 |
| Education (spouse) | 2.02 | 0.98 | 9,731,360 |
| Age at first birth | 20.65 | 3.11 | 11,766,586 |
| Difference in first two kids boys vs girls | 0.02 | 0.71 | 11,766,586 |
| Year | 1991 | 10.32 | 11,766,586 |
| *Panel B: Individual level variables (weighted by sampling weights)* | | | |
| Had more than two kids | 0.60 | 0.49 | 11,760,688 |
| Economically active | 0.51 | 0.50 | 10,269,926 |
| First two children are same sex | 0.51 | 0.50 | 11,760,688 |
| Age | 30.03 | 3.58 | 11,760,688 |
| Education (own) | 1.72 | 0.84 | 11,289,167 |
| Education (spouse) | 1.96 | 0.91 | 9,726,444 |
| Age at first birth | 20.65 | 2.99 | 11,760,688 |
| Difference in first two kids boys vs girls | 0.04 | 0.71 | 11,760,688 |
| Year | 1990 | 9.75 | 11,760,688 |
| *Panel C: Country level variables* | | | |
| Real GDP per capita | 9682 | 9579 | 141 |
| Mean educational attainment | 1.92 | 0.55 | 135 |
| Mean age | 30.03 | 0.82 | 142 |
| Labor force participation (women with at least two children) | 0.43 | 0.24 | 128 |
| Sex imbalance (first two children) | 0.02 | 0.02 | 142 |
| Year | 1989 | 11.77 | 142 |
| *Panel D: Dyadic differences between country pairs* | | | |
| Education (own) | 0.65 | 0.48 | 14,641 |
| Education (spouse) | 0.60 | 0.44 | 14,641 |
| Age | 1.01 | 0.76 | 14,641 |
| Year | 13.07 | 10.29 | 14,641 |
| Real GDP per capita | 10432 | 9464 | 14,400 |
| Sex imbalance (first two children) | 0.015 | 0.014 | 14,641 |
| Total fertilty rate | 0.760 | 0.610 | 14,641 |
| Labor force participation (women with at least two children) | 0.24 | 0.61 | 14,641 |
| Geographic distance (km) | 7847 | 4720 | 14,641 |

Notes: Source: Authors' calculations based on data from the Integrated Public Use Microdata Series-International (IPUMS-I). Observations vary due to missing data.

*children* and *Economically active*), and control for age of mother, own education, and spouse's education, subject to the sample restrictions discussed above. The country-year treatment effects are summarized in Appendix Table 1. Effects are measured in terms the changes in the probability of having more kids and being economically active.

**5. Graphically characterizing heterogeneity**

To motivate our analysis, we start by providing some descriptive figures that help to understand the heterogeneity of the treatment effects in our data. Figure 1 is a funnel plot, which is a scatter plot of the treatment effect of *Same-Sex* on *Had more children* in our sample of 142 complete-data country-year samples against the standard error of the treatment effect. The region within the dotted lines in the figure should contain 95% of the points in the absence of treatment-effect heterogeneity. Figure 1 clearly shows that there is substantial heterogeneity for this treatment effect that goes beyond what one would expect to see were it a homogenous treatment effect with mean-zero random variation. A similar, but less stark, picture arises in Figure 2, which presents the funnel plot of *Same-Sex* on *Economically active* in the 128 samples that have census information on this labor market outcome.

Figures 1 and 2 also highlight the fact that not all country-year treatment effects are statistically significantly different from zero. In Figure 1, approximately three fourths of treatment effects are significant at the 10 per cent level (and two thirds at the 5 per cent level). In Figure 2, approximately one tenth of the treatment effects are significant at standard levels. The differences in significance are driven both by heterogeneity in estimated effects as well as variation in the estimated standard errors. Given the

Figure 1: Funnel Plot of *Same-Sex* and *Having more children*

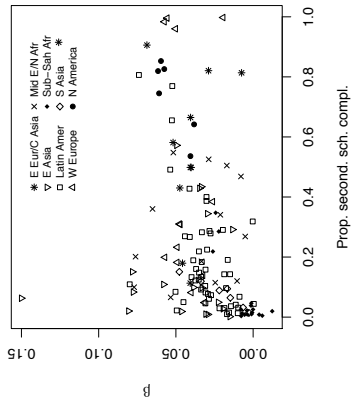

Notes: The funnel plot in this figure is based on data from 142 census samples. Source: Authors' calculations based on data from the *Integrated Public Use Microdata Series -International (IPUMS-I)*.

Figure 2: Funnel Plot of *Same-Sex* and *Being economically active*

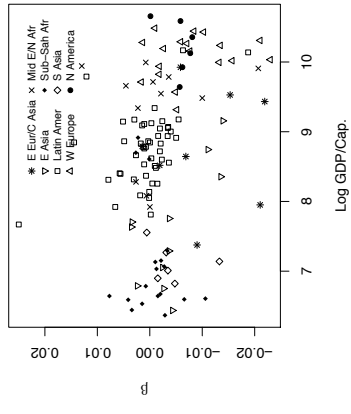

Notes: The funnel plot in this figure is based on data from 128 census samples. Source: Authors' calculations based on data from the *Integrated Public Use Microdata Series -International (IPUMS-I)*.

Figure 3: Treatment effect heterogeneity of *Same-Sex* on *Having more children* by the proportion of women with a completed secondary education.

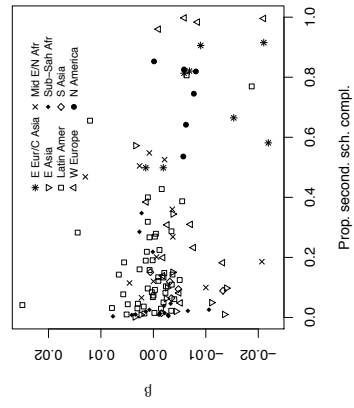

Notes: The graph plots the size of the treatment effect of *Same-Sex* on *Having more children* by the proportion of women with a completed secondary education based on data from 142 census samples. The graph also displays heterogeneity by geographic region. Pearson's correlation: 0.38 (p < 0.001). Source: Authors' calculations based on data from the *Integrated Public Use Microdata Series -International (IPUMS-I)*.

Figure 4: Treatment effect heterogeneity of *Same-Sex* on *Being economically active* by the proportion of women with a completed secondary education.

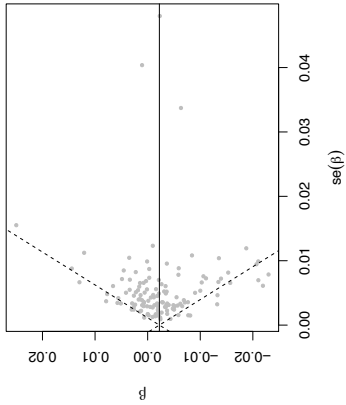

Notes: The graph plots the size of the treatment effect of *Same-Sex* on *Being economically active* by the proportion of women with a completed secondary education based on data from 142 census samples. The graph also displays heterogeneity by geographic region. Pearson's correlation: −0.33 (p<0.001). Source: Authors' calculations based on data from the *Integrated Public Use Microdata Series -International (IPUMS-I)*.

Figure 5: Treatment effect heterogeneity of *Same-Sex* on *Having more children* by log GDP per capita

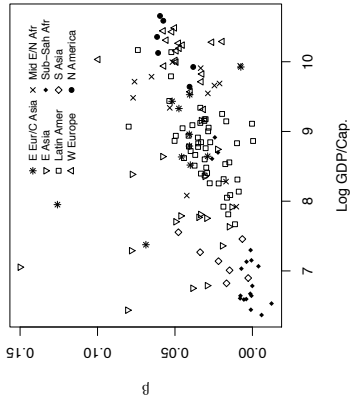

Notes: The graph plots the size of the treatment effect of *Same-Sex* on *Having more children* by log GDP per capita based on data from 142 census samples. The graph also displays heterogeneity by geographic region. Pearson's correlation: 0.39 (p < 0.001). Source: Authors' calculations based on data from the *Integrated Public Use Microdata Series -International (IPUMS-I)*.

Figure 6: Treatment effect heterogeneity of *Same-Sex* on *Being economically active* by log GDP per capita

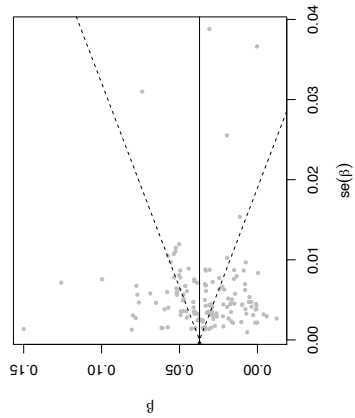

Notes: The graph plots the size of the treatment effect of *Same-Sex* on *Being economically active* by log GDP per capita based on data from 142 census samples. The graph also displays heterogeneity by geographic region. Pearson's correlation: 0.26 (p = 0.004). Source: Authors' calculations based on data from the *Integrated Public Use Microdata Series -International (IPUMS-I)*.

substantial heterogeneity in the precision of our estimates, in our subsequent analysis, we weight the country-year treatment effects by the standard error of the treatment effect.

The next set of figures investigates whether any of the treatment effect heterogeneity documented in Figures 1 and 2 is correlated with heterogeneity in observable covariates. In Figures 3 and 4 we plot the size of the treatment effect of *Same-Sex* on *Had more children* (Figure 3) and *Economically active* (Figure 4) on the y-axis against the proportion of women with a completed secondary education based on data from 142 census samples (on the x-axis). Figure 3 shows a positive relationship that suggests that the treatment effect is larger in countries with a higher proportion of educated mothers. The same figure also displays heterogeneity based on geographic region, indicating small (or zero) effects in countries of Sub-Saharan Africa. The corresponding effects for *Economically active* in Figure 4 are suggestive of a negative relationship between the treatment effect size and the level of education in a country, without a strong geographical pattern.

Finally, in Figures 5 and 6 we repeat the analysis from the previous two figures but instead we describe the heterogeneity with respect to log GDP per capita in a country. Figure 5 shows a striking linear pattern, suggesting the treatment effects of *Same-Sex* on *Had more children* increase with income per capita. Since the proportion of women with a secondary education and the log of GDP per capita are clearly correlated, it implies that Figures 3-6 are not informative of the relative importance of one covariate over another. Nonetheless, these graphs as well as the funnel plots presented earlier all provide suggestive evidence showing that there is substantive heterogeneity for both of our treatment effects and that this heterogeneity is associated with levels of development.

## 6. Homogeneity tests

The next step in our analysis is to quantify the heterogeneity depicted in Figures 1 and 2, and to establish that it is statistically significant. We start by presenting, in Table 2, the results of Cochran's Q tests for effect homogeneity (Cochran, 1954), which quantify what is depicted in Figures 1 and 2 in terms of the heterogeneity in the observed effect sizes against what one would obtain as a result of sampling error if there were a homogenous effect. The resulting test statistics, which are tested against the Chi-square distribution with degrees of freedom equal to the number of effects minus one, are extremely large (and the resulting p-values are essentially zero) and confirm statistically the visual impression of treatment effect heterogeneity for both treatment effects from Figures 1 and 2. The results are similar when the unit of observation is the country-year-education group.

Given that there is heterogeneity, for the second test we investigate if the effects are distributed in a manner that resemble a normal distribution. For this we have implemented an inverse-variance weighted Shapiro-Francia (wSF) test for normality of effect estimates. This test modifies the Shapiro-Francia test for normality (Royston 1993) by taking into account the fact that the country-year treatment effects are estimated with different levels of precision. Our modification involves using an inverse-variance weighted correlation coefficient as the test statistic rather than the simple sample correlation coefficient. The test statistic is the squared correlation between the sample order statistics and the expected values of normal distribution order statistics. In our specific example, where the outcome is *Had more children*, we take the order sample values for our 142 country-year observations and look at the squared correlation between the ordered statistics from our sample and the expected ordered percentiles of the

Table 2: Heterogeneity tests

| Outcome | Effect specification | N* | Q-test statistic** (p-value) | wSF-test statistic*** (p-value) |
|---|---|---|---|---|
| More kids | Country-year | 142 | 13,998 (<.0001) | 0.9345 (<.0001) |
| | Country-year-ed. category | 533 | 15,573 (<.0001) | 0.9433 (<.0001) |
| Economically active | Country-year | 128 | 224.26 (<.0001) | 0.948 -0.0002 |
| | Country-year-ed. category | 477 | 586.26 (<.0001) | 0.8592 (<.0001) |

Notes: *Number of studies, which varies over the two outcomes because of incomplete data over available samples for the economically active indicator.
**Q test of effect homogeneity. Degrees of freedom are 141 for More kids and 127 for Economically active.
***Inverse-variance weighted Shapiro-Francia (wSF) test for normality of effect estimates. The test statistic is the squared correlation between the sample order statistics and the expected values of normal distribution order statistics.

standard normal distribution. The results in Table 2 confirm that for both of our outcome variables we can reject that the correlation is 1, i.e., we can reject the hypothesis of unconditional normality. This result is not surprising in light of the visual evidence presented in Figures 1 and 2, which suggested that the distribution of our country-year effects is over-dispersed from what a normal distribution would look like.

The rejection of homogeneity suggests the need to use available covariates to extrapolate to new contexts. In our example, the set of covariates is limited. At the micro level we have only the basic demographic characteristics included in the standardized IPUMS data, but a somewhat larger set of country-year covariates. We expect that such limits to available covariates would be typical of experimental evidence bases. With a limited set of covariates, we can remain agnostic about what covariates to include and just incorporate all of them into a flexible specification without encountering degrees of freedom problems, using the LASSO regularization to prune interactions and higher order terms.

**7. Characterizing external validity: the external validity function and unconditional relationships**

In this section, we present a graphical analysis of the importance of context covariates such as education, log GDP per capita, and geographical distance in improving extrapolations of the same-sex treatment effects. We conduct this descriptive analysis using the external validity function, which characterizes how prediction errors from reference to target locations vary as a function of the context-level covariate differences between locations. (This is analogous to the bias function in Heckman, Ichimura, Smith, and Todd 1998.)

Specifically, we extrapolate the treatment effect to a target context adjusting only for unit-level covariates from the reference context(s) (age, education, etc. from Table 1, Panel A). For descriptive transparency, we use only a single reference context, although in practice using the full set of available reference contexts is more efficient. This yields a prediction error estimate, $\hat{\zeta}_{ci}$, for each target context $c$ from reference context $i$. We then evaluate how this prediction error varies in $V_{ck} - \bar{V}_{ik}$, where $\bar{V}_{ik}$ is the mean of the $k$th context level covariate from the reference context(s) $i$ used to generate the prediction for site $c$. For a single reference context $i$, $\bar{V}_{ik}$ is simply the value of the $k$-th context level covariate, but for some examples below, we construct a context-level covariate by taking the mean of unit-level covariates. In Figures 7 to 10, we present local linear regressions of prediction error for all reference-target dyads, $\hat{\zeta}_{ci}$, on within-dyad covariate differences $V_{ck} - \bar{V}_{ik}$.

Unconditional external validity function estimates for education are presented in Figure 7. Three features are notable. Prediction error is approximately zero at zero education distance, which is consistent with and provides a test of the unconfounded location assumption. Prediction error increases with increasing differences in education levels; for a one standard deviation education difference (approximately one point on the four-point scale) error increases by approximately 0.1 (relative to the world treatment effect of 0.04 in Figure 1). The figure also plots +/- two standard errors of the external validity function, which is relatively flat over the range of -2 to +2 educational differences, but increases at greater differences.

Figure 8 shows a similar pattern when we explore how the prediction error changes with GDP per capita. The error at zero GDP per capita distance is close to zero, and increases to about 0.1 for a one standard deviation GDP per capita difference

Figure 7: Unconditional external validity function: local linear regression of prediction error on standardized differences in education

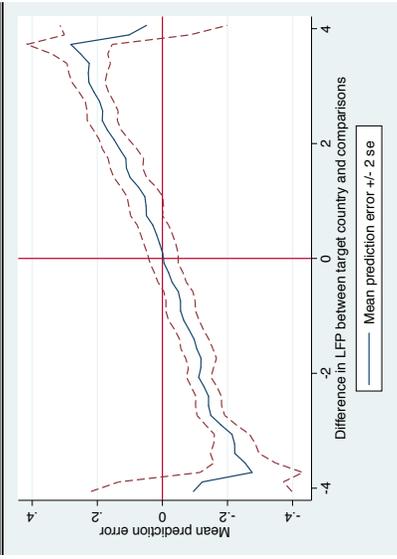

Notes: The graph plots the local polynomial regression of the dyadic prediction error against the standardized education difference between target and comparison country, where the education difference is standardized by its standard deviation (0.83). The variables are further described in Table 1. Source: Authors' calculations based on data from the Integrated Public Use Microdata Series-International (IPUMS-I).

Figure 8: Unconditional external validity function: local linear regression of prediction error on standardized differences in log GDP per capita

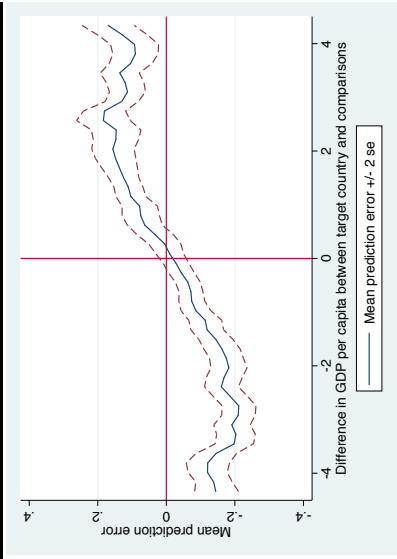

Notes: The graph plots the local polynomial regression of the dyadic prediction error against the standardized GDP difference between target and comparison country, where the GDP difference is standardized by its standard deviation (59680). The variables are further described in Table 1. Source: Authors' calculations based on data from the Integrated Public Use Microdata Series-International (IPUMS-I).

Figure 9: Unconditional external validity function: local linear regression of prediction error on standardized differences in women's labor force participation

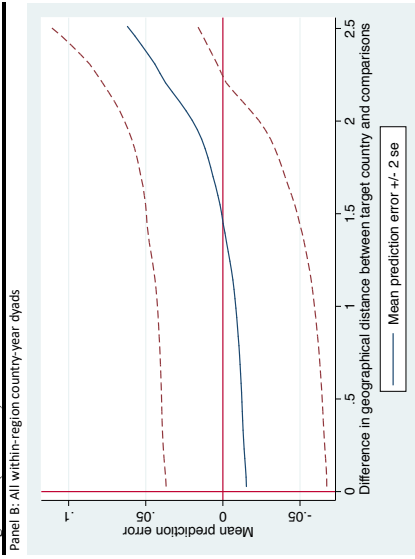

Notes: The graph plots the local polynomial regression of the dyadic prediction error against the standardized labor force participation difference between target and comparison country, where the labor force participation difference is standardized by its standard deviation (0.21). The variables are further described in Table 1. Source: Authors' calculations based on data from the Integrated Public Use Microdata Series-International (IPUMS-I).

Figure 10: Unconditional external validity function: local linear regression of prediction error on standardized geographical distance

Panel A: All country-year dyads

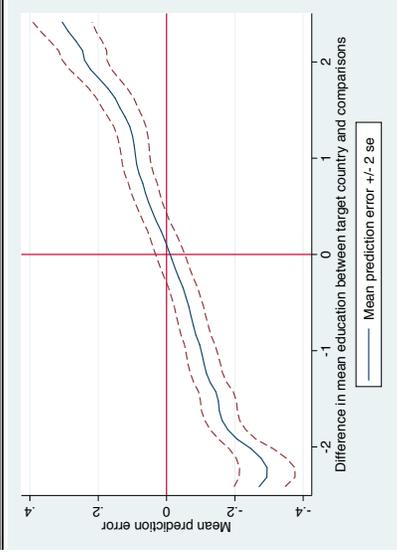

Notes: The graph plots the local polynomial regression of the dyadic prediction error against the standardized geographical distance between target and comparison country, where the geographical distance is standardized by its standard deviation (4800 km). The variables are further described in Table 1. Source: Authors' calculations based on data from the Integrated Public Use Microdata Series-International (IPUMS-I).

Figure 10 (continued)

Panel B: All within-region country-year dyads

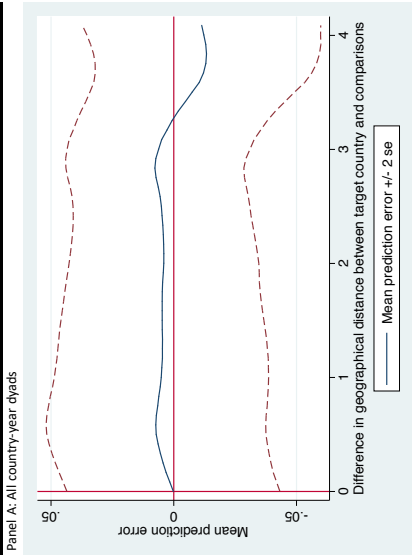

Notes: The graph plots the local polynomial regression of the dyadic prediction error against the standardized geographical distance between target and comparison country, for within-region dyads (where regions defined as North and South America, Europe, Asia, and Africa) and with the geographical distance is standardized by its standard deviation (4800 km). The variables are further described in Table 1. Source: Authors' calculations based on data from the Integrated Public Use Microdata Series-International (IPUMS-I).

(approximately $10,000). In Figure 9 we focus on women's labor force participation differences and again we observe that any deviations in labor force participation distance are associated with higher prediction error.

In Figure 10, we present external validity function estimates with respect to geographic distance, measured as the standardized distance in kilometers between the centroid of a target and comparison country (where a one standard deviation difference is approximately 4800 km). Geographic distance is presumed to proxy for various cultural, climactic, or other geographically clustered sources of variation in fertility. Looking across all country-years, in Panel A of Figure 10, we do not find a significant relationship between geographical distance and prediction error. Non-linear features of geographical distance, most notably oceans, complicate this relationship. To account for this, in Panel B of Figure 10, we present differences within contiguous regions (North and South America, Europe, Asia, and Africa). Again, we do not find any statistically significant relationship for distances less than 10,000 km. The estimated external validity function is positively sloped: for distances in excess of approximately 10,000 km, there is a statistically significant increase in prediction error.[3]

## 8. Characterizing external validity: conditional relationship

In this section we continue our characterization of external validity by estimating the multivariate relationship between prediction error and the full range of dyadic covariate differences. The goal of this analysis is twofold. First, it allows us to test with a precise standard error the validity of the unconfounded location assumption. Second, it gives us a descriptive sense of which context covariates are most important when extrapolating.

---

[3] Appendix figures 1-3 present results of tests for the unconfounded location assumption in the spirit of the tests used by Hotz et al. (2005) and Gechter (2015). They are analogous to Figures 7-10, but instead extrapolate the $Y_i(0)$ distribution. The graphs pass through the origin which is what we would expect if unconfounded location holds.

We regress $\hat{\zeta}_{ci}$, for each target context $c$ - reference context $i$ dyad on $V_{ck} - \bar{V}_{ik}$, for $k=1,\ldots,K$, the within-dyad covariate differences, where we adjust the standard errors using the Cameron and Miller (2014) dyadic cluster-robust estimator.

The results from this exercise are presented in Tables 3 and 4, where we standardize covariate differences. In order to interpret the coefficients it is useful to note that the standard deviation of the education variable is close to 0.5, for age it is about 0.75 years, for census year it is 10 years, for log GDP per capita is about 9,464 dollars, and for distance it is about 4700 km.

In columns (1) to (9) of Table 3, we run the prediction error regressions one covariate at a time, giving us prediction error linear regressions corresponding to Figures 7-10. Most covariates (measured as standard deviations of reference-target differences in education, education of spouse, age of the mother, year of census, log GDP per capita and labor force participation) are statistically significant, with a one standard deviation covariate difference increasing prediction error by 0.05 to 0.1, an order of magnitude approximately between one and two times the treatment effect (with differences in mother's age and total fertility rate leading to even larger errors). Geographical distance notably is not statistically significant.

In columns (10) to (11) of Table 3, we estimate multivariate prediction error regressions. Five main observations can be drawn from the results. First, the constant in the regressions is close in magnitude to, and not statistically significantly different from, zero, matching the finding from Figures 7-10 that when covariate differences between the reference and target location are small prediction error is also small. This is consistent

Table 3: Extrapolation prediction error regressions for *Having more children* - with covariates

| Standardized Difference between country pairs in: | Prediction error (1) | Prediction error (2) | Prediction error (3) | Prediction error (4) | Prediction error (5) | Prediction error (6) | Prediction error (7) | Prediction error (8) | Prediction error (9) | Prediction error (10) | Prediction error Excluding sex selectors¶ (11) |
|---|---|---|---|---|---|---|---|---|---|---|---|
| Education of mother (σ=0.48) | 0.0693*** (0.0108) | | | | | | | | | -0.000160 (0.0245) | 0.00686 (0.0260) |
| Education of father (σ=0.44) | | 0.0648*** (0.0110) | | | | | | | | -0.0159 (0.0229) | -0.0210 (0.0250) |
| Age of mother (σ=0.76) | | | 0.114*** (0.00831) | | | | | | | 0.0676*** (0.0124) | 0.0666*** (0.0130) |
| Census year (σ=10.3) | | | | 0.0545*** (0.0116) | | | | | | 0.00491 (0.00765) | 0.00492 (0.00793) |
| log GDP per capita (σ=9464) | | | | | 0.0830*** (0.0133) | | | | | -0.0187 (0.0137) | -0.0174 (0.0149) |
| Sex ratio imbalance (σ=0.02) | | | | | | 0.0117 (0.0142) | | | | 0.0113 (0.00851) | 0.0142 (0.0118) |
| Labor force participaiton (σ=0.17) | | | | | | | 0.0657*** (0.0129) | | | -0.00164 (0.00923) | -0.00428 (0.00973) |
| Total fertiltiy rate (σ=0.61) | | | | | | | | -0.122*** (0.0111) | | -0.0975*** (0.0133) | -0.0972*** (0.0136) |
| Distance in KM (σ=4720) | | | | | | | | | 0.0116 (0.0187) | 0.0107 (0.00864) | 0.0107 (0.00819) |
| Distance squared | | | | | | | | | -0.00400 (0.00538) | -0.00367 (0.00283) | -0.00346 (0.00272) |
| Constant | -0.00333 (0.00523) | 0.00249 (0.00396) | 0.00342 (0.00654) | 0.00521 (0.00721) | -0.00610 (0.00833) | 0.00500 (0.0133) | 0.00199 (0.0108) | 0.00445 (0.00828) | 0.000805 (0.0127) | 0.00212 (0.00749) | 0.00298 (0.00770) |
| Observations | 14,641 | 14,641 | 14,641 | 14,641 | 14,400 | 14,641 | 14,641 | 14,641 | 14,641 | 14,400 | 12,321 |
| R-squared | 0.184 | 0.141 | 0.549 | 0.117 | 0.173 | 0.000 | 0.148 | 0.638 | 0.000 | 0.723 | 0.724 |

Notes: The table shows extrapolation prediction error regressions as described in Section 8 of the paper. The left-hand-side variable is reference-to-target prediction error in the country-year dyad. The right-hand-side variables are standardized referene-to-target differences in covariates, where the standardization is given in parentheses. Standard errors are adjusted for dyadic clustering using Cameron and Miller's (2014) procedure. ¶Column 11 excludes China, India, Vietnam, and Nepal. Source: Authors' calculations based on data from the Integrated Public Use Microdata Series-International (IPUMS-I).

with the unconfounded location (assumption C1). Second, many of the variables are statistically significant, although we note that education and labor force participation lose significance once the other controls are included. Third, the size of the prediction error due to covariate differences is generally large relative to an average treatment effect in the sample of 0.04. Fourth, it is noteworthy that the effects of GDP per capita and total fertility rate are negative in column (10). Since the unconditional effect of GDP per capita differences is positive in column (5), this reflects the counter-intuitive nature of the variation identifying the conditional coefficient: variation in GDP per capita conditional on a similar education, age, and labor force participation profile of women is presumably quite limited. At the same time, the coefficient on the difference in total fertility rate is negative both unconditionally (in column (8)) and conditionally (column (10)). This implies that the treatment effect is decreasing in total fertility rate, so comparing a reference country-year to a target country-year with a lower total fertility leads to negative prediction error (under-estimation of the treatment effect). Fifth, the sex ratio imbalance enters positively, implying that it is indeed important to consider the degree of sex selectivity within countries when extrapolating the treatment effect. This remains true even when we drop the most notable sex-selectors from the sample (China, India, Nepal, and Vietnam, column (11)). Furthermore, dropping sex-selecting countries does not meaningfully change the estimated coefficients on covariate differences.

The results in Table 4 for the effect of *Same-Sex* on *Economically active* are similar in three respects. First, the constant is not statistically significantly different from zero at least when all covariates are included in columns (10) to (11), again consistent with unconfounded location (C1). Second, the magnitude of prediction error generated by reference-covariate target differences is large relative to the treatment effect. Third,

Table 4: Extrapolation prediction error regressions for *Being economically active* - with covariates

| Standardized Difference between country pairs in: | Prediction error (1) | Prediction error (2) | Prediction error (3) | Prediction error (4) | Prediction error (5) | Prediction error (6) | Prediction error (7) | Prediction error (8) | Prediction error (9) | Prediction error (10) | Prediction error Excluding sex selectors[¶] (11) |
|---|---|---|---|---|---|---|---|---|---|---|---|
| Education of mother ($\sigma$=0.48) | -0.0237 (0.0170) | | | | | | | | | 0.0576** (0.0270) | 0.0535** (0.0268) |
| Education of father ($\sigma$=0.44) | | -0.0178 (0.0160) | | | | | | | | -0.0413* (0.0251) | -0.0348 (0.0238) |
| Age of mother ($\sigma$=0.76) | | | -0.0688*** (0.0177) | | | | | | | -0.00789 (0.0125) | -0.00468 (0.0127) |
| Census year ($\sigma$=10.3) | | | | -0.0760*** (0.0154) | | | | | | -0.0113 (0.00938) | -0.0110 (0.00970) |
| log GDP per capita ($\sigma$=9464) | | | | | -0.0143 (0.0176) | | | | | 0.0427*** (0.0129) | 0.0359** (0.0140) |
| Sex ratio imbalance ($\sigma$=0.02) | | | | | | 0.0372* (0.0197) | | | | -0.00125 (0.00823) | 0.0698 (0.0111) |
| Labor force participation ($\sigma$=0.17) | | | | | | | -0.152*** (0.00884) | | | -0.167*** (0.00789) | -0.166*** (0.00819) |
| Total fertility rate ($\sigma$=0.61) | | | | | | | | 0.0571*** (0.0189) | | -0.00839 (0.0144) | -0.00943 (0.0140) |
| Distance in KM ($\sigma$=4720) | | | | | | | | | 0.0983*** (0.0308) | 0.0292* (0.0157) | 0.0379** (0.0172) |
| Distance squared | | | | | | | | | -0.0120*** (0.00442) | -0.00606* (0.00314) | -0.00890** (0.00368) |
| Constant | 0.0748** (0.0348) | 0.0761** (0.0375) | 0.0536** (0.0255) | 0.0672** (0.0293) | 0.0784** (0.0367) | 0.0776** (0.0376) | 0.0255* (0.0132) | 0.0603** (0.0269) | -0.0337 (0.0263) | 0.00335 (0.0159) | 0.00198 (0.0161) |
| Observations | 14,641 | 14,641 | 14,641 | 14,641 | 14,400 | 14,641 | 14,641 | 14,641 | 14,641 | 14,400 | 12,321 |
| R-squared | 0.006 | 0.001 | 0.069 | 0.160 | 0.001 | 0.022 | 0.735 | 0.086 | 0.031 | 0.825 | 0.816 |

Notes: The table shows extrapolation prediction error regressions as described in Section 8 of the paper. The left-hand-side variable is reference-to-target prediction error in the country-year dyad. The right-hand-side variables are standardized reference-to-target differences in covariates, where the standardization is given in parentheses. Standard errors are adjusted for dyadic clustering using Cameron and Miller's (2014) procedure. [¶]Column 11 excludes China, India, Vietnam, and Nepal. Source: Authors' calculations based on data from the Integrated Public Use Microdata Series-International (IPUMS-I).

covariate differences enter both positively (sex ratio imbalance, total fertility rate) and negatively (age of the mother, calendar year, and labor force participation of women) both unconditionally and conditional on other covariates. This reflects different patterns of treatment effect heterogeneity: a positive coefficient on the reference-target covariate difference implies that the treatment effect is increasing in the covariate (so if the target country has a higher value of the covariate, one overestimates the treatment effect in the reference country), a negative coefficient the opposite.

While the results in Tables 3 and 4 allow us to compare the simultaneous importance of a range of covariates difference on prediction error, they do not allow us to judge the importance of micro vs. country-level covariates. Since dyads are formed at the country-year level, micro-level covariates differences are aggregated to that level. In order to get at this issue, we perform the following exercise for each country-year sample. We take a given country-year as the target country, and all of the other country-years are treated as reference sites. Pooling the data from the reference sites, we run a separate regression for the treated and the control observations, and we use these to predict the treatment and the control outcomes and the treatment effect in the target site. We consider four cases in terms of possible sets of regressors: (1) one without any covariates, which recovers the unadjusted estimates; (2) the individual micro covariates including age of the mother, a set of dummies on mother's educational attainment, a set of dummies on the education of the spouse, age at first marriage, as well as all the possible interactions of these individual-level variables; (3) macro covariates consisting of log GDP per capita, labor force participation, dummies for British and French legal origin, as well as a variables for the latitude and longitude of a country; and (4) the combined covariates that consist of the union of micro (group 2) and macro variables (group 3). We calculate prediction error in the same manner as above.

This exercise generates a data point for each country-year with a separate prediction from each of the four covariate sets. We plot the distribution of these prediction errors for *Had more children* in Figure 11 and for *Economically active* in Figure 12. The four groups are unadjusted (solid line), micro variables only (wide dashed line), macro variables only (small dashed lines), and micro and macro variables together (dotted line). In panel A of each figure, we plot the density estimates of these prediction errors, while in panel B we plot the CDFs of the absolute prediction error.

Looking at Figure 11, we observe that in the case of *Had more children*, both micro and macro variables contribute in pushing prediction error towards zero, dominating the scenario of no covariates. In the density plots, inclusion of covariates brings in the tails toward zero, and in the CDF plot the error distribution is drawn toward zero. However, the contribution of the macro variables is much stronger and considerably reduces the error. The results in Figure 12, which use *Economically active* as the outcome variable of interest, provide an even starker picture. In this case, micro variables do not seem useful in terms of reducing the prediction error, a finding that is in line with the arguments provided in Pritchet and Sandefur (2013). But equally remarkable is how well macro variables do in terms of reducing prediction error. The implication of these results is that a set of easily available cross-country variables has the potential to be useful in analyzing external validity. This also raises concerns about generating extrapolations solely on the basis of micro-level data, an issue that Hotz et al. (2005), Stuart et al. (2011), and Hartman et al. (2015) were unable to investigate due to the limitations of their evidence bases.

Figure 11: Individual versus macro covariates for *Having more children*

Panel A: Density estimate - prediction error

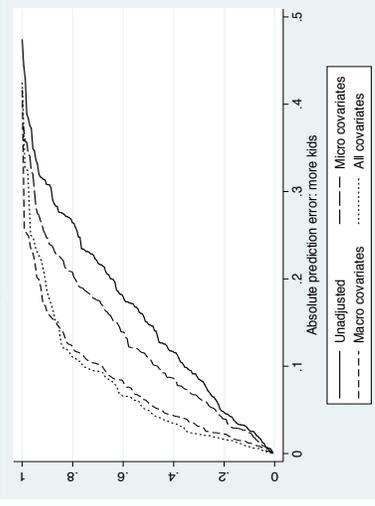

Figure 11 (continued)

Panel B: CDF - absolute prediction error

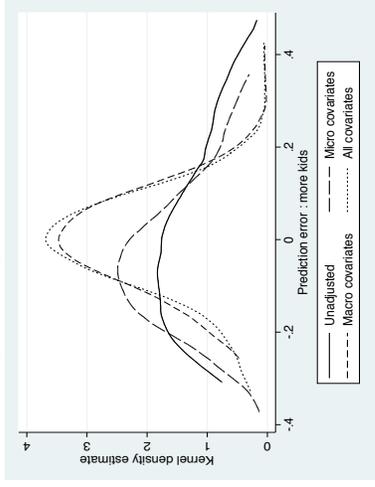

Notes: The graph plots the density estimates of the prediction error and CDF of the absolute prediction error based on the procedure described in Section 9 of the paper. Source: Authors' calculations based on data from the Integrated Public Use Microdata Series-International (IPUMS-I).

Figure 12: Individual versus macro covariates for *Being economically active*

Panel A: Density estimate - prediction error

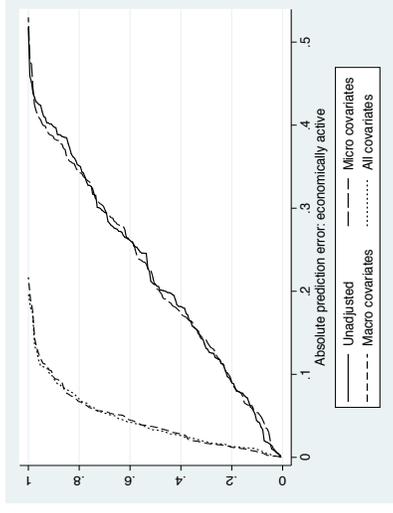

Figure 12 (continued)

Panel B: CDF - absolute prediction error

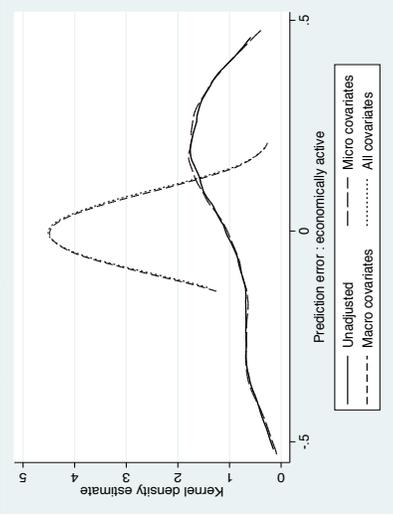

Notes: The graph plots the density estimates of the prediction error and CDF of the absolute prediction error based on the procedure described in Section 9 of the paper. Source: Authors' calculations based on data from the Integrated Public Use Microdata Series-International (IPUMS-I).

Finally, we obtain similar results on the importance of context-level covariates when we use the LASSO regularization to specify the approximating functions characterized in expression (7) above. This allows us to evaluate the importance of interactions and higher order terms. In the application in section 10.2 below, we use these LASSO-pruned models to generate predictions. Appendix Figure 4 shows the solution paths for the interaction terms in the series expansion. The solution path reveals that an error-minimizing specification (in terms of Mallow's Cp-statistic) is quite sparse in the interaction terms retained. Moreover, macro-level and macro-micro interaction terms dominate the LASSO solution paths, which means that they are the variables that LASSO selects to produce a parsimonious and error-minimizing specification. Even in the fully saturated specification, the macro and macro-micro interaction terms that we have included dominate in terms of explanatory power (evident in looking at the standardized coefficient values displayed all the way to the right in the graphs of the full LASSO solution paths, Panels A and B). These results confirm two impressions arising from the exploratory analysis above: first, much of the effect heterogeneity is attributable to macro-level variation, and second, to the extent that micro-level variables matter in explaining effect heterogeneity, the influence of these micro-level variables is strongly moderated by macro-level moderation (e.g., the age of mothers moderates treatment effects, but in a manner that differs depending on macro context).

## 9. The accumulation of evidence and out-of-sample prediction error

Our results so far imply that with sufficient covariate data, we can extrapolate the treatment effect with zero prediction error on average, when the reference and target contexts are similar, particularly with respect to context covariates. We now consider if and how the accumulation of experiments over time improves our ability to extrapolate to new settings or alternatively how well we are able to extrapolate with only a small experimental evidence base. The results, by year, are plotted in Figures 13 and 14 for our two outcomes.

For the target country-years observed in a given year, *t*, we extrapolate the treatment effect and estimate prediction error using the reference sample available in years *t*-1 and earlier. We restrict the reference sample and generate extrapolations in four different ways: (method 1, small dashed lines line) pooling all country-years available up to year *t*-1, we extrapolate using the average treatment effect in the pooled reference sample; (method 2, solid line) we extrapolate using the treatment effect from the lowest prediction error reference country-year as selected by the prediction-error model (from Table 3) fit to data up to year *t*-1; (method 3, dotted line) we extrapolate using the treatment effect from the nearest country-year by geographical distance excluding own-country comparisons; and (method 4, wide dashed lines line) we extrapolate using the treatment effect from the nearest country-year by geographic distance, allowing own-country comparisons.

A number of interesting patterns arise from this exercise. First, consider the comparison of pooling all available country-years (method 1, in small dashed lines) versus the best reference country-year selected by the model (method 2, in solid line). The results confirm that when using our model we get much lower prediction error compared to pooling all the samples available. Second, the pattern of prediction error over time from using the model-selected reference country-year (method 2) shows that the accumulation of more samples plays a modest but meaningful role in reducing the prediction error. The role of adding samples is modest in the sense that the prediction

Figure 13: Prediction error with different comparison groups of *Same-Sex* on *Having more children*

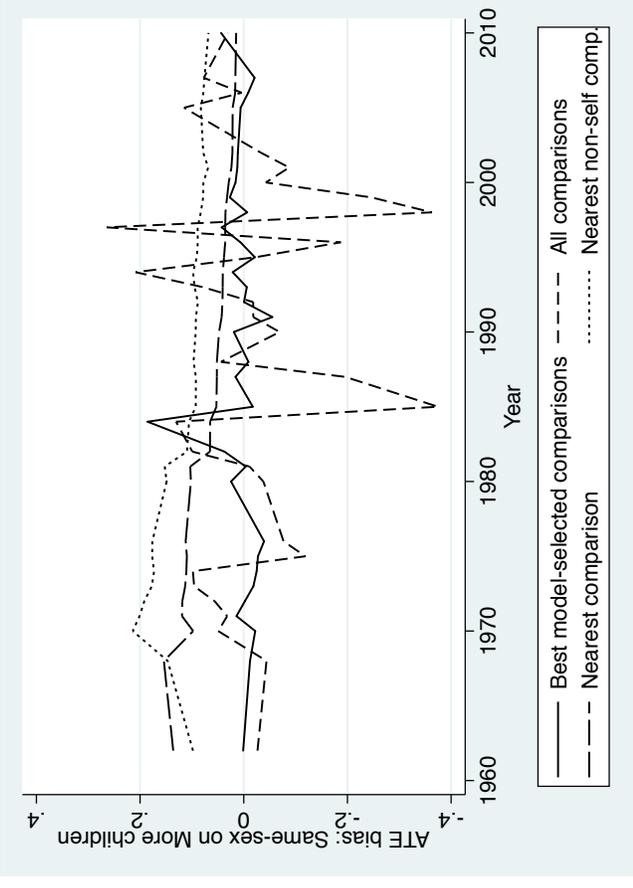

Notes: The graph plots the prediction error for target country-years available up to the year on the x-axis using the procedure described in Section 9 of the paper and four groups of reference countries: (1) all the available country years, (graphed as the blue line), (2) the best comparison country-year as predicted by our model (graphed as the red line), (3) the nearest country-year by distance excluding own-country comparisons (graphed as the orange line), and (4) the nearest country-year by distance, allowing own-country year comparisons. The variable on the X-axis refers to the year when a census was taken. The variables are further described in Table 1. Source: Authors' calculations based on data from the *Integrated Public Use Microdata Series - International (IPUMS-I)*.

Figure 14: Prediction error with different comparison groups of *Same-Sex* on *Being economically active*

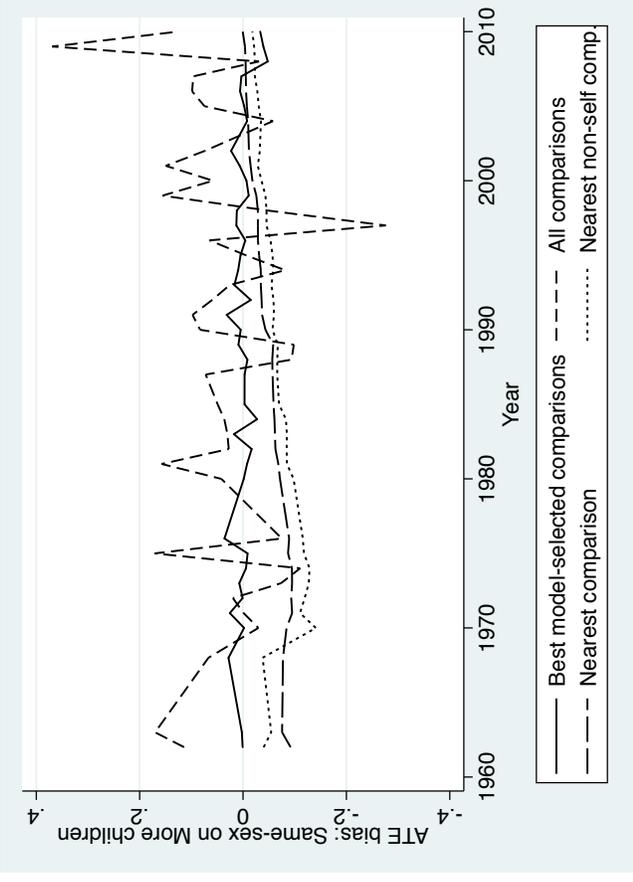

Notes: The graph plots the prediction error for target country-years available up to the year on the x-axis using the procedure described in Section 9 of the paper and four groups of reference countries: (1) all the available country years, (graphed as the blue line), (2) the best comparison country-year as predicted by our model (graphed as the orange line), and (4) the nearest country-year by distance excluding own-country comparisons (graphed as the orange line), and (4) the nearest country-year by distance, allowing own-country year comparisons. The variable on the X-axis refers to the year when a census was taken. The variables are further described in Table 1. Source: Authors' calculations based on data from the Integrated Public Use Microdata Series-International (IPUMS-I).

error from the model-selected reference country-year hovers between 0.08 and -0.05. This suggests that the model is reasonably accurate in making predictions even with a limited number of available samples. But adding samples is also meaningful in the sense that the prediction error tightens considerably (ranging between 0.02 and -0.03) from 1985 onward.

We can also compare the model-based approach to simple rule-of-thumb selection criteria. First is the rule of thumb of choosing the nearest country-year by geographic distance, but excluding own-country samples (method 3, dotted line), and second is the same geographic rule of thumb, but allowing for own-country samples from previous years (method 4, wide dashed line). In both cases, the prediction error becomes smaller over time, likely because the geographically-nearest match tends to be quite similar. We see marked improvements from allowing own-country reference samples from previous years, suggesting that cross-sectional heterogeneity is important. We also find that neither rule of thumb tends to perform as well as the model-based approach, particularly when available reference samples are sparse.

Overall, we draw three conclusions from this analysis. First, without a sufficient number of experiments extrapolating the treatment effect is challenging; while the model-informed approach (method 2) performs well on average, in our data, its reliability is sensitive to year-to-year variation in the reference sample until around 1985 (by which point we have accumulated 54 country-year samples). Second, with a sufficiently large evidence base, rules of thumb are somewhat reliable. Third, in both rich and sparse data environment the model informed approach tends to dominate either pooled estimation or the simple rules of thumb.

## 10. Applications

While the natural experiment we have examined, the effect of *Same-sex* on fertility, clearly is not a intervention that could or would be implemented by a policy maker, as a thought experiment we treat it as such, and in this section examine how our framework would be used to address two questions a policy maker could face: (1) where to locate an experiment to minimize average prediction error over a set of target sites, and (2) when to rely on extrapolation from an existing experimental evidence base rather than running a new experiment in a target site of interest.

*10.1 Where to locate an experiment*

Imagine a policy researcher interested in characterizing how the effect of an intervention varies around the world as in Imbens (2010, p. 420) or Rubin (1992), but with limited resources to implement new experiments. In this section we examine what the evidence base implies for the best location of new experimental sites given the goal of generating evidence that generalizes globally.

At the country-year level, our regressions above suggest that prediction error should be low for locations with low covariate distance to the evidence base. In assessing such covariate distance, the question is how to weight different covariates. With knowledge of the estimates in Tables 3 and 4 (column (10) in each table) one would weight each covariate by its conditional importance for external validity, or more directly one could also weight each covariate by its conditional influence on the country-year treatment effect. Figure 15 provides confirmation for this intuition. We use each country-year to predict the other country-years in our sample, where the x-axis plots each country-year by the percentile of its composite covariate, i.e., the sum of covariates

Figure 15: Mean prediction error on percentile of comparison country composite treatment-effect predictor, using one site to predict all others

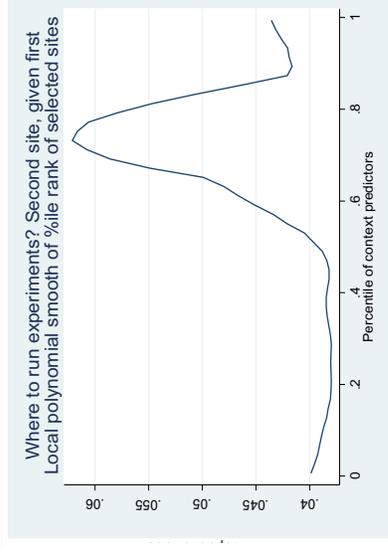

Notes: On the x-axis each country-year is ranked based on its percentile of a composite treatment effect predictor. The composite predictor is a weighted average country-year covariates weighted by their effect on the country-year treatment effect. The y-axis show the mean prediction error from using the site on the x-axis to predict all other country-years. Source: Authors' calculations based on data from the Integrated Public Use Microdata Series-International (IPUMS-I).

Figure 16: Mean prediction error on average Mahalanobis distance of the comparison country-year to all target country-years

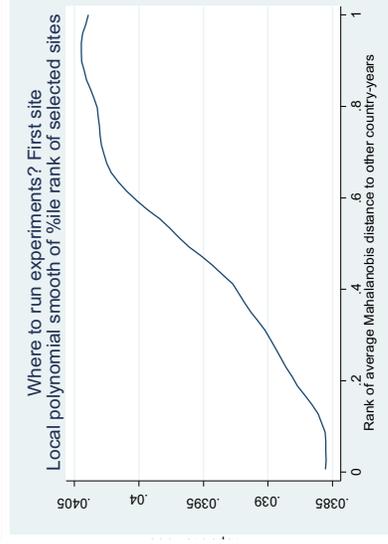

Notes: On the x-axis each country-year is ranked based on its avearge Mahalanobis distance to all other country-years. The y-axis show the mean prediction error from using the site on the x-axis to predict all other country-years. Source: Authors' calculations based on data from the Integrated Public Use Microdata Series-International (IPUMS-I).

Figure 17: Mean prediction error, given the first comparison site, on percentile of composite treatment-effect predictor covariate, using two sites to predict the others

Figure 18: Mean prediction error, given the first comparison site, on average Mahalanobis distance of the comparison country-year to all target country-years, using two sites to predict others

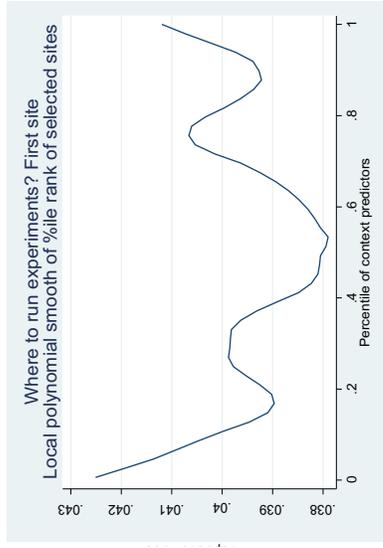

Notes: On the x-axis each country-year is ranked based on its avearge Mahalanobis distance to all other country-years. The y-axis show the mean prediction error from using the site on the x-axis in addition to the first selected comparison site to predict all other country-years. Source: Authors' calculations based on data from the Integrated Public Use Microdata Series-International (IPUMS-I).

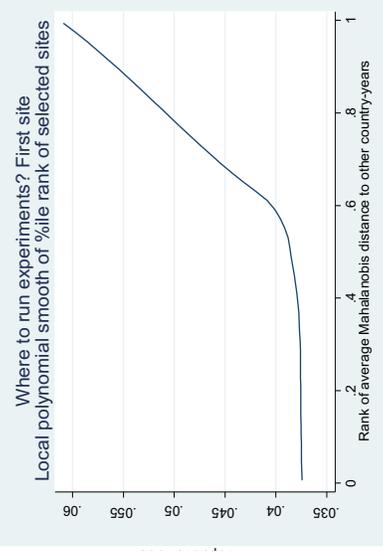

Notes: On the x-axis each country-year is ranked based on its percentile of a composite treatment effect predictor. The composite predictor is a weighted average country-year covariates weighted by their effect on the country-year treatment effect. The y-axis show the mean prediction error from using the site on the x-axis to predict all other country-years. Source: Authors' calculations based on data from the Integrated Public Use Microdata Series-International (IPUMS-I).

Figure 19: To experiment or extrapolate? A graphical illustration of the decision problem

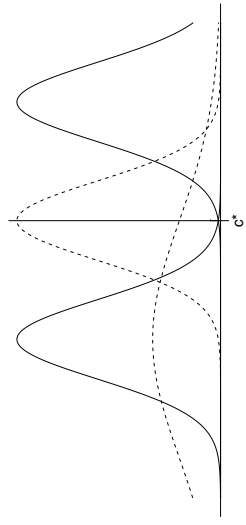

Notes: Solid line = experiment not warranted. Dashed line = experiment warranted.

weighted by their conditional predictive relevance for the treatment effect, and where the y-axis plots the associated mean error from predicting the treatment effect for other country-years. We see immediately that the lowest average prediction error is indeed at the median, which turns out to be the United States in 1980.

The challenge in thinking of this prescriptively is that a policy maker will not know the conditional importance of each covariate for external validity without first running the full set of experiments. In Figure 16, we consider an alternative that does not rely on knowledge of the treatment effect; namely, we compute the average covariate Mahalanobis distance between each country-year and the other country-years. The covariate Mahalanobis distance accounts for redundancy due to correlations between covariates. It therefore accounts for all of the information in the linear external validity function specification that we can obtain without knowing the regression coefficients. The figure plots average prediction error against the rank of average distance of a country-year from other country-years. Again, it is evident that the country-year with the lowest average distance to other country-years offers the lowest prediction error of the treatment effect; the relationship is also monotonic. Carrying the thought experiment further, in Figures 17 and 18 we consider adding a second country-year, conditional on the first choice. As such, in these figures, we are using two countries to make predictions. Again, the lowest prediction error is associated with country-years that are in the middle of the covariate distribution or that have the lowest average covariate distance to other country-years.

If one had to choose only a single site to locate an experiment in order to learn about a collection of sites, the results show that choosing in a manner that minimizes Mahalanobis distance would offer an estimate that extrapolates with expected prediction error that is low relative to alternative sites. If, however, the goal is to add new experiments to an existing evidence base so as to characterize how effects vary, then these results recommend selecting sites that maximize Mahalanobis distance in the covariates as specified in the external validity function. It is for such sites that the evidence base is unreliable in predicting treatment effects. These results are similar in spirit, though different in details, than those of Stuart et al. (2011) and Tipton (2014). Stuart et al. (2011) use the standardized mean difference of experimental site-selection propensity scores to summarize how a selected and unselected sites differ in their respective covariate distributions. Tipton (2014) uses the Bhattacharyya distance between the propensity scores distributions of the selected and unselected sites. As the literature on matching has demonstrated, the optimal distance metric will depend on the underlying covariate and outcome distributions (Abadie and Imbens 2006).

*10.2 To experiment or to extrapolate?*

Now suppose a policy maker wants to make an evidence-based policy decision of whether or not to implement a program. The policy maker has a choice between using the existing evidence base versus generating new evidence by carrying out an experiment with the target population. That being the case, the choice is really between whether the existing evidence base can provide a reliable enough estimate of what would be found from the new experiment, thus making the new experiment unnecessary. Essentially, the policy-maker would want to work with predictions that use available micro-level covariate data to account for differences in micro-level population characteristics and available macro-level covariate data to account for differences in macro-level context characteristics. As the previous section anticipated, we should expect the reliability of such predictions to depend on the amount of covariate data available and also how irregular are the covariate values for the target context as compared to what is contained in the evidence base.

One might imagine different ways to characterize the loss function governing this decision. We develop an approach based on the assumption that a new experiment is only worthwhile if the existing evidence base is sufficiently ambiguous about the potential effects of the treatment for the target population. Formally, this means that the policy maker will decide that the existing evidence is sufficient to determine policy if a 95% prediction interval surrounding the conditional mean prediction for the target site is entirely on one or another side of some critical threshold, $c^*$. We also assume the experiment that the policy maker could run with the target population is adequately well powered that she would find it worthwhile to run the experiment if the existing evidence is ambiguous. Figure 19 illustrates the decision problem graphically. If the predictive interval resembles either of the solid-line distributions, then the evidence is certain enough to rule out the need for an experiment. If the interval resembles either of the dashed line distributions, then the existing evidence is too vague and a new experiment is warranted.

This is a reduced-form characterization of any number of more fully-fledged analyses. A fully Bayesian decision analysis under a Normal model could begin with the premise that the policy maker implements the program if the posterior distribution for the program effect provides a specified degree of certainty that the effect will be above some minimal desirable effect value. Then, $c^*$ and the relevant prediction interval could be defined as a function of the minimum desirable effect value, the level of certainty required, posterior variance, and the moments of the predictive distribution. With $c^*$ and the relevant prediction interval defined, the analysis would otherwise proceed as we describe here.

Recall that in expression (5) we defined $\hat{\tau}_1$ as the estimator for the target population treatment effect and in expression (6) we defined the error $\epsilon_{(1)}$ that characterizes how $\tau_{(1)}$, the treatment effect for any given target population, differs from the expected target population effect, $\tau_1$. We apply a working assumption that $\epsilon_{(1)}$ is normally distributed with mean zero and variance $\sigma_\zeta^2$. This is a substantive assumption on the distribution of treatment effects. We generate a prediction $\hat{\tau}_1$, and apply another working assumption that $\hat{\tau}_1$ is normally distributed with mean $\tau_1$ and estimation variance $\sigma_1^2$. This assumption can be taken to approximate the large sample distribution of $\hat{\tau}_1$ as a consistent, linear estimator. Note that $\tau_{(1)}$ and $\hat{\tau}_1$ are statistically independent by virtue of the assumed process through which individuals are assigned to the target population (assumption C1) and the fact that outcomes from the target context are not used to estimate $\hat{\tau}_1$. Consider the difference $X = \tau_{(1)} - \hat{\tau}_1$. This difference is a linear combination of independent normal variables, and thus is normal with mean zero and variance $\sigma_\zeta^2 + \sigma_1^2$. Applying the usual parametric results for out-of-sample prediction intervals, a 95% prediction interval for $\tau_{(1)}$ is given by

(8) $\quad PI_1 = \hat{\tau}_1 \pm t_{.025}(\sigma_\zeta^2 + \sigma_1^2),$

where $t_{.025}$ is the appropriate .025 quantile value for the normalized conditional distribution of $\hat{\tau}_1$. The first variance component captures the intrinsic variability of context-level treatment effects, and does not diminish in the number of reference contexts. The second variance component captures estimation variability and goes to zero in the number of reference contexts. The solution to the decision problem is to experiment if $c^* \in PI_1$, and accept the existing evidence otherwise. Under the normality assumptions, this would imply an error rate of 5%.

We estimate the total variance ($\sigma_\zeta^2 + \sigma_1^2$) in a manner that accounts for potential dependency between this variance and covariates. We do so using a leave-one-out approach, similar to that of Orr et al. (2017). We first generate predictions, $\hat{\tau}_{1c}$, for each of the reference contexts in the evidence base, and then we take the difference $\hat{\zeta}_c = \hat{\tau}_{1c} - \hat{\tau}_c$, where $\hat{\tau}_c$ is the effect estimated using the natural experiment in context $c$. We then model the $\log(\hat{\zeta}_c^2)$ values in terms of $(V_c, \overline{W}_c)$ using a series specification analogous to what we used to model the conditional mean. We take the exponentiated predicted value at $(V_{(1)}, \overline{W}_{(1)})$, as our estimate of the total variance.

Figure 20 shows the results of applying this approach to estimating the effects of *Same-sex* on *More kids*. Panel A shows how the cumulative reference sample evolves over time, eventually reaching our 142 complete-data country-year samples and about 10 million observations. Panel B shows the prediction intervals for target country-year (gray bars), arrayed by year. We also plot the actual effect estimates from those country-year samples (black dots) as a way to check on the accuracy of the procedure. The figure shows that the predictive intervals are informative, in that they do not span an extreme range, and they almost always cover the in-sample effect. The intervals become a bit tighter as the evidence base grows over time, although they do not collapse to zero. As a result, even for a decision rule based on a critical value of 0 ($c^* = 0$) and even with over 100 reference samples, the analysis would indicate the need for further experimentation.

That the intervals do not collapse to zero is expected because of the intrinsic variability, and this highlights the crucial role of covariate data for analyses that depend

Figure 20: To experiment or extrapolate? Sample, prediction intervals, and uncertainty estimates

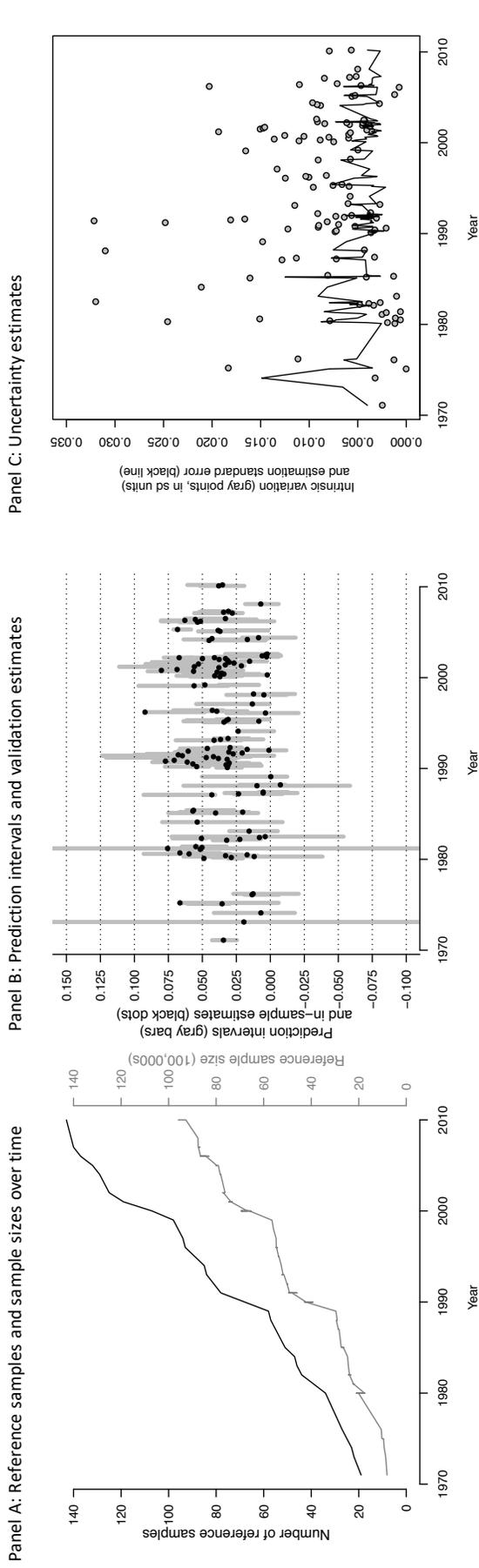

Notes: Panel A shows how the cumulative reference sample evidence base is growing over time in terms of the number of reference country-year samples (black) and the number of reference sample observations (gray). Panel B shows the estimated prediction interval for the effect of *Same-sex* on *More kids* for each target country-year (gray bars) and then, for validation, the actual effect estimates from those country-year samples (black dots). Panel C shows the estimation standard error for each target country-year (black line) and then the estimated intrinsic variation, that is, the estimated standard deviation of the effect distribution at the point in the covariate space for the target country-year. Source: Authors' calculations based on data from the Integrated Public Use Microdata Series-International (IPUMS-I).

on external validity. Unlike the standard error of prediction, the intrinsic variability does not depend on the sample size in a strict sense. Rather, it is a function of the amount of variation left unexplained by the covariates, which remains fixed in this application. Panel C demonstrates this point clearly. The black line traces out the standard error of prediction), which tends toward zero as the reference samples accumulate. The gray dots show the estimates of the intrinsic variation, expressed in standard deviation units and thus on the same scale as the effect estimates. The intrinsic variation always dominates the standard error of prediction, and it remains quite large (relative to the size of the treatment effects) even as the sample size gets huge.

To tighten the intervals further, one would need to reduce the intrinsic variation. This would require either collecting more covariate data or finding ways to better use existing covariates to characterize the conditional effect distribution. Thus, even if rich covariate data are not needed for internal validity, this application shows the crucial role of covariate data in informing decisions that rely on external validity.

## 11. Conclusion

This paper has examined whether, in the context of a specific natural experiment and a data context, it is possible to reach externally valid conclusions regarding a target setting of interest using an evidence base from a reference context. We view this paper as having made six contributions to the literature. First, we provide and implement a simple framework to consider external validity. Second, we come up with a context in which it is possible, and meaningful, to ask and potentially to answer questions of external validity. While randomized and quasi-experiments are run and estimated globally, to our knowledge there is no one design that has been run in as many countries, years, and geographical settings as the *Same-Sex* natural experiment. While it has challenges as a natural experiment, we view our exercise as a possibility result: is external validity – notwithstanding the challenges – possible? Third, we present results that directly answer the central question of external validity, namely the extent to which valid conclusions about a target population of interest can be drawn from the available data. Fourth, we show that, given the accumulation of sufficient evidence, it is possible to draw externally valid conclusions from our evidence base, but the ability to do so is meaningfully improved (over rule of thumb alternatives) by the modeling approach we adopt. Fifth, we show that prediction error can, in general, depend on both individual and context covariates, although for our application, macro-level context covariates dominate. Finally, we considered two applications for our approach. This first showed that experiments located near the middle of the covariate distribution tend to provide the most robust external predictions and that selecting on the maximum covariate Mahalanobis distance is contributes to learning about effect variability efficiently. The second that in some contexts it is possible that a policy maker may choose to extrapolate the treatment effect from an existing experimental evidence base rather than run a new experiment, but that this depends crucially on the richness of available covariate data.

Prescriptively, we would draw four conclusions from our analysis about extrapolating experimental or quasi-experimental evidence from one setting to another. First, the reference and target setting must be similar along economically relevant dimensions, and particularly in terms of macro level features. In our analysis reference-target covariate differences of half a standard deviation created prediction error on the order of the treatment effect. Second, a sufficiently large experimental evidence base is needed for reliable extrapolation; for our data, at least fifty country-year samples were needed before out-of-sample extrapolation became reliable. Third, given sufficient data, accounting for treatment effect heterogeneity in the evidence base is essential in extrapolating the treatment effect. Fourth, modeling treatment effect heterogeneity is important when extrapolating treatment effects in sparse data environments; in data-rich settings, rules of thumb might be sufficient.

While our conclusions are cautiously optimistic, it is important to underline both the caution and the inductive nature of our exercise. Our conclusions are circumscribed by the data and application we have considered. Nonetheless, given the importance of the question and paucity of evidence, we believe even a single attempt to assess the external validity of experimental evidence is valuable, despite its flaws and limitations. A better understanding of our ability to learn from the rapidly accumulating evidence from randomized experiments and quasi-experiments, and to answer key policy and economic questions of interest, will require further extensions and replications of the exercise we have begun here.

## Appendix Table 1: Treatment effects and standard errors by country-year

| Country | Year of census | Treatment effect for Having more kids | Standard error for Having more kids | Treatment effect for Economically active | Standard error for Economically active |
|---|---|---|---|---|---|
| Argentina | 1970 | 0.0495 | 0.0078 | -0.0034 | 0.0061 |
| Argentina | 1980 | 0.0451 | 0.0028 | -0.0019 | 0.0024 |
| Argentina | 1991 | 0.0352 | 0.0023 | -0.0050 | 0.0024 |
| Argentina | 2001 | 0.0283 | 0.0026 | -0.0034 | 0.0028 |
| Armenia | 2001 | 0.1259 | 0.0071 | -0.0210 | 0.0070 |
| Austria | 1971 | 0.0330 | 0.0061 | 0.0016 | 0.0060 |
| Austria | 1981 | 0.0499 | 0.0063 | -0.0157 | 0.0066 |
| Austria | 1991 | 0.0452 | 0.0061 | -0.0135 | 0.0067 |
| Austria | 2001 | 0.0520 | 0.0064 | -0.0103 | 0.0066 |
| Belarus | 1999 | 0.0289 | 0.0041 | -0.0069 | 0.0039 |
| Bolivia | 1976 | 0.0143 | 0.0058 | 0.0018 | 0.0054 |
| Bolivia | 1992 | 0.0187 | 0.0052 | 0.0066 | 0.0061 |
| Bolivia | 2001 | 0.0164 | 0.0050 | 0.0001 | 0.0056 |
| Brazil | 1960 | 0.0156 | 0.0021 | -0.0002 | 0.0015 |
| Brazil | 1970 | 0.0218 | 0.0017 | -0.0015 | 0.0014 |
| Brazil | 1980 | 0.0272 | 0.0018 | 0.0000 | 0.0017 |
| Brazil | 1991 | 0.0399 | 0.0015 | -0.0019 | 0.0015 |
| Brazil | 2000 | 0.0350 | 0.0015 | -0.0021 | 0.0015 |
| Cambodia | 1998 | 0.0292 | 0.0035 | 0.0024 | 0.0032 |
| Chile | 1970 | 0.0293 | 0.0044 | 0.0057 | 0.0034 |
| Chile | 1982 | 0.0308 | 0.0044 | 0.0007 | 0.0036 |
| Chile | 1992 | 0.0410 | 0.0041 | 0.0007 | 0.0034 |
| Chile | 2002 | 0.0302 | 0.0044 | -0.0055 | 0.0043 |
| China | 1982 | 0.0806 | 0.0013 | -0.0043 | 0.0010 |
| China | 1990 | 0.1501 | 0.0014 | -0.0024 | 0.0009 |
| Colombia | 1973 | 0.0186 | 0.0027 | 0.0030 | 0.0024 |
| Colombia | 1985 | 0.0374 | 0.0027 | -0.0010 | 0.0027 |
| Colombia | 1993 | 0.0369 | 0.0025 | 0.0026 | 0.0024 |
| Colombia | 2005 | 0.0351 | 0.0025 | 0.0012 | 0.0022 |
| Costa Rica | 1973 | -0.0004 | 0.0084 | 0.0049 | 0.0072 |
| Costa Rica | 1984 | 0.0503 | 0.0081 | 0.0003 | 0.0068 |
| Costa Rica | 2000 | 0.0388 | 0.0071 | 0.0015 | 0.0065 |
| Cuba | 2002 | 0.0412 | 0.0047 | -0.0016 | 0.0056 |
| Ecuador | 1974 | 0.0097 | 0.0044 | 0.0079 | 0.0037 |
| Ecuador | 1982 | 0.0175 | 0.0044 | 0.0011 | 0.0039 |
| Ecuador | 1990 | 0.0332 | 0.0043 | 0.0056 | 0.0042 |
| Ecuador | 2001 | 0.0298 | 0.0043 | -0.0012 | 0.0043 |

Appendix Table 1 continued: Treatment effects and standard errors by country-year

| Country | Year of census | Treatment effect for Having more kids | Standard error for Having more kids | Treatment effect for Economically active | Standard error for Economically active |
|---|---|---|---|---|---|
| Egypt | 1996 | 0.0424 | 0.0014 | 0.0006 | 0.0012 |
| France | 1962 | 0.0316 | 0.0035 | -0.0052 | 0.0029 |
| France | 1968 | 0.0401 | 0.0035 | -0.0049 | 0.0031 |
| France | 1975 | 0.0325 | 0.0034 | -0.0018 | 0.0034 |
| France | 1982 | 0.0492 | 0.0031 | -0.0132 | 0.0032 |
| France | 1990 | 0.0492 | 0.0035 | -0.0077 | 0.0036 |
| France | 1999 | 0.0477 | 0.0036 | -0.0071 | 0.0035 |
| Ghana | 2000 | 0.0012 | 0.0038 | -0.0033 | 0.0028 |
| Greece | 1971 | 0.0770 | 0.0056 | -0.0100 | 0.0053 |
| Greece | 1981 | 0.0761 | 0.0047 | -0.0006 | 0.0043 |
| Greece | 1991 | 0.0651 | 0.0046 | -0.0036 | 0.0052 |
| Greece | 2001 | 0.0517 | 0.0056 | 0.0008 | 0.0066 |
| Guinea | 1983 | 0.0079 | 0.0070 | -0.0106 | 0.0076 |
| Guinea | 1996 | 0.0055 | 0.0047 | 0.0041 | 0.0050 |
| Hungary | 1970 | 0.0407 | 0.0074 | NA | NA |
| Hungary | 1980 | 0.0475 | 0.0057 | NA | NA |
| Hungary | 1990 | 0.0518 | 0.0059 | -0.0219 | 0.0061 |
| Hungary | 2001 | 0.0405 | 0.0075 | -0.0153 | 0.0082 |
| India | 1983 | 0.0148 | 0.0045 | -0.0035 | 0.0049 |
| India | 1987 | 0.0219 | 0.0044 | -0.0133 | 0.0046 |
| India | 1993 | 0.0337 | 0.0050 | -0.0031 | 0.0052 |
| India | 1999 | 0.0478 | 0.0049 | 0.0006 | 0.0050 |
| Iraq | 1997 | 0.0104 | 0.0022 | 0.0000 | 0.0017 |
| Israel | 1972 | 0.0288 | 0.0072 | -0.0021 | 0.0070 |
| Israel | 1983 | 0.0212 | 0.0063 | NA | NA |
| Israel | 1995 | 0.0079 | 0.0062 | 0.0130 | 0.0067 |
| Italy | 2001 | 0.0262 | 0.0033 | 0.0013 | 0.0046 |
| Jordan | 2004 | 0.0170 | 0.0046 | 0.0026 | 0.0046 |
| Kenya | 1989 | -0.0039 | 0.0032 | -0.0028 | 0.0036 |
| Kenya | 1999 | 0.0071 | 0.0033 | -0.0013 | 0.0031 |
| Kyrgyz Republic | 1999 | 0.0688 | 0.0058 | -0.0090 | 0.0050 |
| Malaysia | 1970 | 0.0150 | 0.0076 | 0.0035 | 0.0105 |
| Malaysia | 1980 | 0.0307 | 0.0088 | -0.0135 | 0.0104 |
| Malaysia | 1991 | 0.0226 | 0.0066 | -0.0111 | 0.0073 |
| Malaysia | 2000 | 0.0331 | 0.0068 | -0.0139 | 0.0072 |
| Mali | 1987 | 0.0010 | 0.0045 | 0.0034 | 0.0055 |
| Mali | 1998 | 0.0077 | 0.0038 | 0.0077 | 0.0048 |

# Appendix Table 1 continued: Treatment effects and standard errors by country-year

| Country | Year of census | Treatment effect for Having more kids | Standard error for Having more kids | Treatment effect for Economically active | Standard error for Economically active |
|---|---|---|---|---|---|
| Mexico | 1970 | 0.0100 | 0.0043 | 0.0017 | 0.0040 |
| Mexico | 1990 | 0.0310 | 0.0014 | -0.0024 | 0.0012 |
| Mexico | 1995 | 0.0337 | 0.0069 | -0.0003 | 0.0074 |
| Mexico | 2000 | 0.0337 | 0.0013 | -0.0009 | 0.0013 |
| Mongolia | 1989 | 0.0133 | 0.0080 | NA | NA |
| Mongolia | 2000 | 0.0495 | 0.0087 | 0.0034 | 0.0071 |
| Nepal | 2001 | 0.0167 | 0.0023 | -0.0048 | 0.0025 |
| Pakistan | 1973 | 0.0027 | 0.0031 | -0.0015 | 0.0016 |
| Pakistan | 1998 | 0.0065 | 0.0010 | NA | NA |
| Palestine | 1997 | 0.0051 | 0.0053 | -0.0037 | 0.0034 |
| Panama | 1960 | 0.0113 | 0.0154 | 0.0250 | 0.0155 |
| Panama | 1970 | 0.0088 | 0.0087 | 0.0001 | 0.0099 |
| Panama | 1980 | 0.0149 | 0.0087 | -0.0036 | 0.0096 |
| Panama | 1990 | 0.0442 | 0.0089 | -0.0003 | 0.0087 |
| Panama | 2000 | 0.0332 | 0.0086 | 0.0144 | 0.0088 |
| Peru | 1993 | 0.0276 | 0.0030 | -0.0005 | 0.0028 |
| Peru | 2007 | 0.0302 | 0.0031 | 0.0010 | 0.0031 |
| Philippines | 1990 | 0.0296 | 0.0015 | -0.0037 | 0.0017 |
| Philippines | 1995 | 0.0347 | 0.0015 | NA | NA |
| Philippines | 2000 | 0.0335 | 0.0016 | NA | NA |
| Portugal | 1981 | 0.0534 | 0.0078 | 0.0023 | 0.0082 |
| Portugal | 1991 | 0.0245 | 0.0077 | 0.0046 | 0.0085 |
| Portugal | 2001 | 0.0334 | 0.0088 | -0.0207 | 0.0094 |
| Puerto Rico | 1970 | 0.0196 | 0.0255 | NA | NA |
| Puerto Rico | 1980 | 0.0537 | 0.0107 | NA | NA |
| Puerto Rico | 1990 | 0.0526 | 0.0111 | 0.0121 | 0.0112 |
| Puerto Rico | 2000 | 0.0523 | 0.0115 | -0.0188 | 0.0119 |
| Puerto Rico | 2005 | 0.0739 | 0.0310 | -0.0063 | 0.0337 |
| Romania | 1977 | 0.0457 | 0.0036 | NA | NA |
| Romania | 1992 | 0.0401 | 0.0032 | -0.0019 | 0.0030 |
| Romania | 2002 | 0.0407 | 0.0036 | 0.0014 | 0.0040 |
| Rwanda | 1991 | 0.0006 | 0.0038 | -0.0015 | 0.0017 |
| Rwanda | 2002 | 0.0040 | 0.0047 | -0.0066 | 0.0029 |
| Saint Lucia | 1980 | 0.0308 | 0.0388 | -0.0023 | 0.0480 |
| Saint Lucia | 1991 | 0.0003 | 0.0366 | 0.0011 | 0.0404 |
| Senegal | 1988 | 0.0038 | 0.0041 | -0.0010 | 0.0047 |
| Senegal | 2002 | 0.0006 | 0.0044 | -0.0021 | 0.0049 |

Appendix Table 1 continued: Treatment effects and standard errors by country-year

| Country | Year of census | Treatment effect for Having more kids | Standard error for Having more kids | Treatment effect for Economically active | Standard error for Economically active |
|---|---|---|---|---|---|
| Slovenia | 2002 | 0.0075 | 0.0097 | -0.0058 | 0.0078 |
| South Africa | 1996 | 0.0261 | 0.0029 | 0.0001 | 0.0029 |
| South Africa | 2001 | 0.0222 | 0.0029 | 0.0027 | 0.0028 |
| South Africa | 2007 | 0.0242 | 0.0063 | 0.0022 | 0.0051 |
| Spain | 1991 | 0.0572 | 0.0040 | -0.0018 | 0.0045 |
| Spain | 2001 | 0.0472 | 0.0051 | -0.0026 | 0.0064 |
| Switzerland | 1970 | 0.0195 | 0.0102 | -0.0059 | 0.0088 |
| Switzerland | 1980 | 0.0557 | 0.0097 | -0.0210 | 0.0099 |
| Switzerland | 1990 | 0.0575 | 0.0105 | -0.0084 | 0.0108 |
| Switzerland | 2000 | 0.0502 | 0.0119 | -0.0010 | 0.0123 |
| Tanzania | 1988 | -0.0123 | 0.0027 | 0.0015 | 0.0021 |
| Tanzania | 2002 | 0.0014 | 0.0022 | -0.0020 | 0.0021 |
| Thailand | 1970 | 0.0195 | 0.0041 | NA | NA |
| Thailand | 1980 | 0.0463 | 0.0067 | NA | NA |
| Thailand | 1990 | 0.0777 | 0.0068 | NA | NA |
| Thailand | 2000 | 0.0580 | 0.0060 | NA | NA |
| Uganda | 1991 | -0.0059 | 0.0029 | -0.0029 | 0.0035 |
| Uganda | 2002 | 0.0001 | 0.0022 | 0.0008 | 0.0029 |
| United Kingdom | 1991 | 0.0996 | 0.0076 | -0.0230 | 0.0079 |
| United States | 1960 | 0.0406 | 0.0034 | -0.0057 | 0.0030 |
| United States | 1970 | 0.0382 | 0.0033 | -0.0062 | 0.0033 |
| United States | 1980 | 0.0609 | 0.0015 | -0.0077 | 0.0015 |
| United States | 1990 | 0.0616 | 0.0015 | -0.0081 | 0.0015 |
| United States | 2000 | 0.0575 | 0.0016 | -0.0059 | 0.0016 |
| United States | 2005 | 0.0597 | 0.0038 | -0.0001 | 0.0038 |
| Venezuela | 1971 | 0.0169 | 0.0035 | 0.0051 | 0.0034 |
| Venezuela | 1981 | 0.0309 | 0.0033 | 0.0029 | 0.0034 |
| Venezuela | 1990 | 0.0286 | 0.0032 | -0.0040 | 0.0031 |
| Venezuela | 2001 | 0.0799 | 0.0031 | -0.0035 | 0.0030 |
| Vietnam | 1989 | 0.0386 | 0.0024 | -0.0027 | 0.0019 |
| Vietnam | 1999 | 0.0782 | 0.0027 | -0.0037 | 0.0023 |

Source: Treatment effect and standard errors by country-year of *Same-Sex* on *Having more children* and *Being economically active*. Source: Authors' calculations based on data from the Integrated Public Use Microdata Series-International (IPUMS-I).

Appendix Figure 1: Testing for unconfounded location: local linear regression of Y(0) prediction error on standardized differences in women's labor force participation

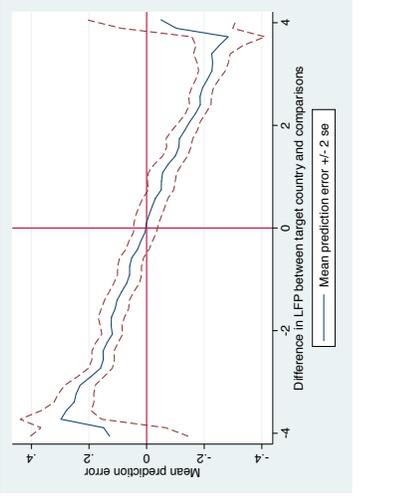

Notes: The graph plots the local polynomial regression of the difference between actual Y(0) and predicted Y(0) against the standardized education difference between target and comparison country, where the education difference is standardized by its standard deviation (0.82). The variables are further described in Table 1. Source: Authors' calculations based on data from the Integrated Public Use Microdata Series-International (IPUMS-I).

Appendix Figure 2: Testing for unconfounded location: local linear regression of Y(0) prediction error on standardized differences in GDP per capita

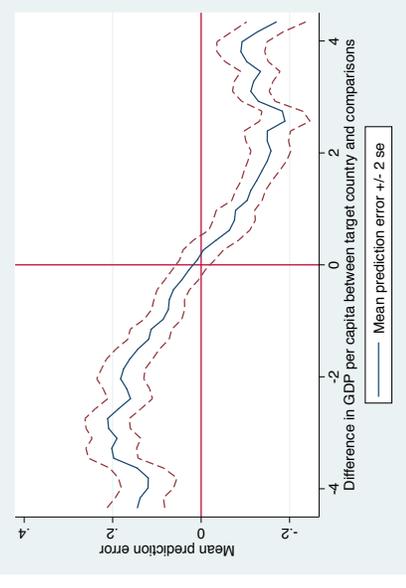

Notes: The graph plots the local polynomial regression of the difference between actual Y(0) and predicted Y(0) against the standardized difference in GDP per capita between target and comparison country, where the education difference is standardized by its standard deviation ($9680). The variables are further described in Table 1. Source: Authors' calculations based on data from the Integrated Public Use Microdata Series-International (IPUMS-I).

Appendix Figure3 : Testing for unconfounded location: local linear regression of Y(0) prediction error on standardized differences in GDP per capita

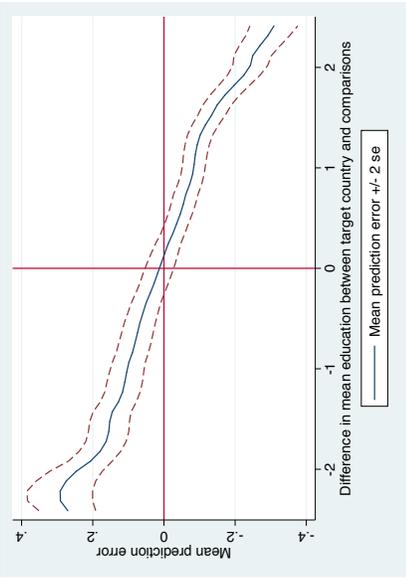

Notes: The graph plots the local polynomial regression of the difference between actual Y(0) and predicted Y(0) against the standardized difference in GDP per capita between target and comparison country, where the education difference is standardized by its standard deviation ($9680). The variables are further described in Table 1. Source: Authors' calculations based on data from the Integrated Public Use Microdata Series-International (IPUMS-I).

Appendix Figure 4: LASSO solution paths for series approximation interaction terms

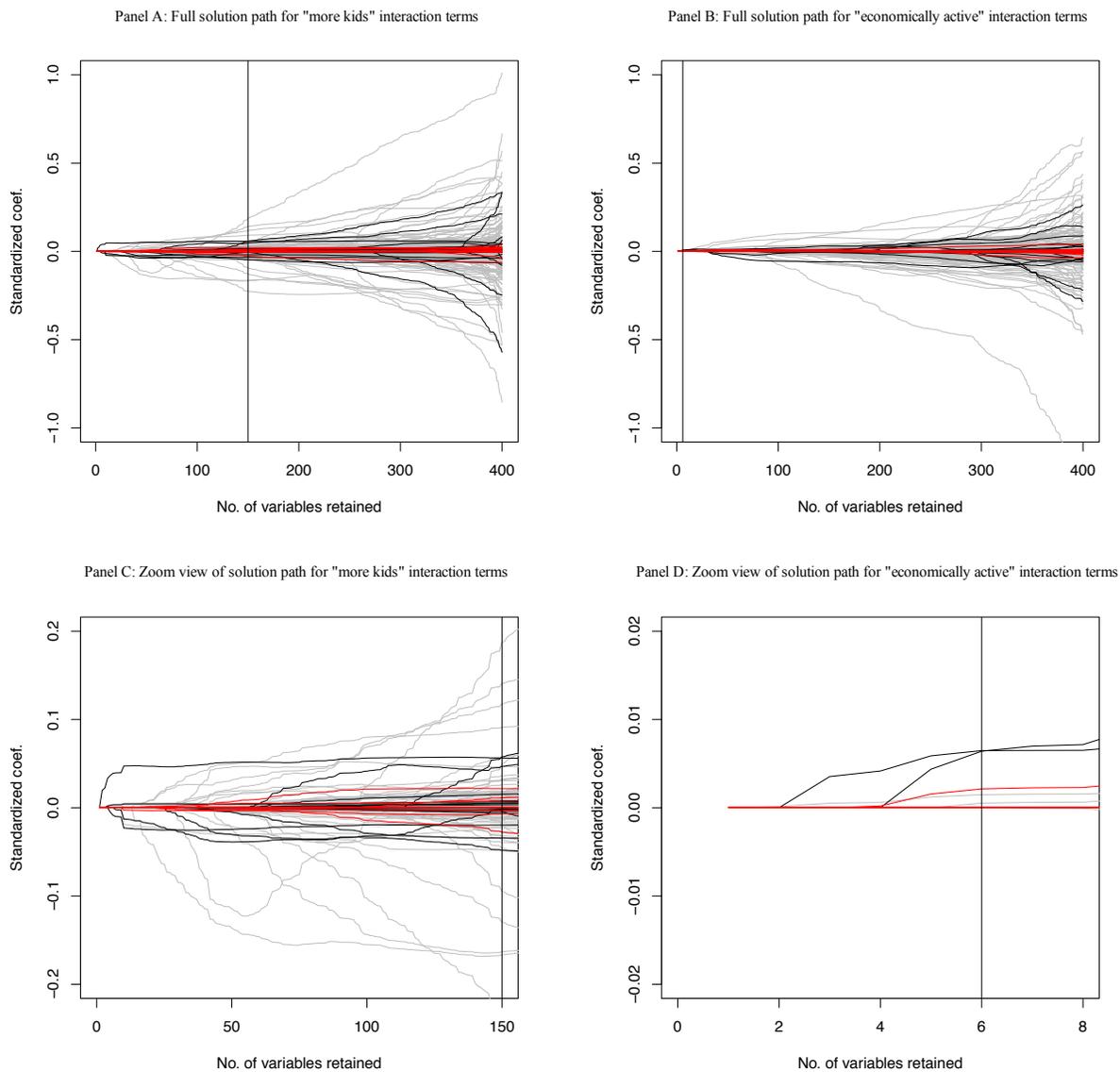

Notes: The graphs plot, on the y-axis, standardized coefficient values for treatment-covariate interaction terms in the series approximation for the more kids (left) and economically active (right) outcomes, and on the x-axis, the number of variables retained under LASSO regularization as one loosens the penalty parameter from including only an intercept (at left in each graph) to including all terms in the series (at right in each graph). The black vertical line shows the point at which the specification minimizes Mallow's $C_p$-statistic. Panels A and B show the full solution path through the full saturated second-order series expansion, while panels C and D zoom to the neighborhood where Cp is minimized. Micro-level covariates are colored red, macro-level covariates are colored black, and macro-micro interactions are colored gray for the lines drawing out the coefficient values in the solution paths. Source: Authors' calculations based on data from the *Integrated Public Use Microdata Series-International (IPUMS-I)*.